\documentclass[reprint, NumberedRefs]{JASA}

\usepackage{color}

\begin{document}

\title[PINN for Acoustic Resonance Analysis in 1D Tube]{Physics-informed neural network for acoustic resonance analysis in a one-dimensional acoustic tube}
\author{Kazuya Yokota}
\email{yokokazu@vos.nagaokaut.ac.jp}
\author{Takahiko Kurahashi}
\affiliation{Department of Mechanical Engineering, Nagaoka University of Technology, Nagaoka, Niigata, 940-2188, Japan}
\author{Masajiro Abe}
\affiliation{Department of System Safety Engineering, Nagaoka University of Technology}

\begin{abstract}
This study devised a physics-informed neural network (PINN) framework to solve the wave equation for acoustic resonance analysis. The proposed analytical model, ResoNet, minimizes the loss function for periodic solutions and conventional PINN loss functions, thereby effectively using the function approximation capability of neural networks while performing resonance analysis. Additionally, it can be easily applied to inverse problems. The resonance in a one-dimensional acoustic tube, and the effectiveness of the proposed method was validated through the forward and inverse analyses of the wave equation with energy-loss terms. In the forward analysis, the applicability of PINN to the resonance problem was evaluated via comparison with the finite-difference method. The inverse analysis, which included identifying the energy loss term in the wave equation and design optimization of the acoustic tube, was performed with good accuracy.

[This article may be downloaded for personal use only. Any other use requires prior permission of the author and AIP Publishing. This article appeared in [The Journal of the Acoustical Society of America, 156(1), 30-43 (2024)] and may be found at \href{https://doi.org/10.1121/10.0026459}{https://doi.org/10.1121/10.0026459}.]
\end{abstract}

\maketitle

\section{\label{sec:1} Introduction}
Machine learning has made significant advances over the last decade, yielding state-of-the-art methods in numerous fields including speech synthesis\cite{Synth-WaveNet, Synth-Tacotron} and image recognition\cite{Image-AlexNet, Image-SqueezeNet}. Advances in machine learning have enhanced numerical simulation technology, which has been applied to turbulence models\cite{SciML-turbulence1, SciML-turbulence2}, design optimization of machine materials\cite{SciML-materials}, and to accelerate the computation time in fluid analysis\cite{SciML-acc1, SciML-acc2}. Recent years have seen an increased demand for machine learning in numerical simulation as it effectively addresses inverse problems such as design optimization and parameter identification.
The physics-informed neural network (PINN), proposed by Raissi et al. \cite{PINN_raissi}, is a numerical analysis method that introduces constraints on the governing partial differential equations (PDEs) in the loss functions of a neural network. In PINN, a loss function based on PDE residuals evaluated by automatic differentiation (AD)\cite{PINN_AD} along with the loss functions of the initial and boundary conditions are defined and then minimized to obtain the simulation results. The PINN can be used for inverse problems to identify uncertain parameters of the governing PDEs. It has also been applied in thermodynamics and fluid mechanics for parameter identification of the equation of state\cite{PINN_eos}, application to inverse problems in supersonic flows\cite{PINN_supersonic}, and identification of the thermal diffusivity of materials\cite{PINN_thermal}.

Despite the aforementioned advances in PINN, their application in acoustics has been limited. Moreover, although the wave equation governing the acoustic phenomena is linear, it is subject to reflection, interference, and diffraction, and its solutions have a wide range of amplitudes and frequencies\cite{WaveEq_complex1}. Therefore, the dynamics of sound waves can be extremely complex, hindering the development of PINN for acoustic analysis \cite{WaveEq_complex2}. Few studies have analyzed acoustic resonance using PINN.

Existing studies that employed PINN for wave equation analyses\cite{WaveEq_complex2,PINN_WaveEq1,PINN_WaveEq2} have considered only unidirectional traveling waves or a small number of acoustic wave reflections and interferences. Thus, they do not apply to resonance phenomena. Several practical PINN based on the wave equation have been reported for seismic wave analysis\cite{PINN_Seismic1,PINN_Seismic2,PINN_Seismic3}; however, they are for large-scale Earth’s ground and inapplicable to sound fields with complex dynamics, such as those of human-scale acoustic instruments. 

In this context, a previous study on acoustic holography using PINN based on the Kirchhoff--Helmholtz integral in the frequency domain\cite{PINN_KH1,PINN_KH2} demonstrated the applicability of PINN to acoustic resonance. N. Borrel-Jensen et al. constructed PINN by introducing reflection boundary conditions and reported that PINN could analyze 1D acoustic resonance using actual physical conditions of air \cite{PINN_Borrel_1D}. They also reported 3D acoustic resonance analysis\cite{PINN_Borrel_3D} using DeepONet\cite{DeepONet} and showed that although it takes several hours to learn, predictions can be made on the millisecond scale. Schmid et al. proposed a method to identify acoustic boundary admittances from noisy sound pressure data using PINN based on the Helmholtz equation in the frequency domain \cite{Schmid}. The results of Schmid et al. demonstrated that the PINN based on frequency-domain equations can perform not only forward analysis but also acoustic inverse analysis with high accuracy. These reports suggest the possibility of acoustic analysis using neural networks.

Time-domain analysis of resonance is critical in the field of acoustics. For example, the two-mass model\cite{Time_2mass1,Time_2mass2} and body-cover model\cite{Time_BodyCover1,Time_BodyCover2} describing vocal fold vibrations in speech production, describe vocal fold vibrations as equations in the time domain. The equations describing lip motion in brass instruments\cite{Time_Brass1,Time_Brass2} and reed motion in woodwinds\cite{Time_Wood1,Time_Wood2} are also described as time-domain equations. These coupled vibration-acoustic analyses often include time-dependent and nonlinear terms such as vocal collisions\cite{Yokota_VocalFolds}, which require time-domain simulations. Therefore, time-domain analysis of resonance using PINN could considerably benefit various acoustics fields. PINN can perform inverse analysis with the same neural network structure as forward analysis and can be used to identify uncertain parameters of the governing PDEs, as mentioned above. Therefore, PINN has the potential to greatly advance the field of inverse analysis in acoustics, such as the diagnosis of speech disorders known as Glottal Inverse Filtering (GIF)\cite{GIF}, the estimation of vocal tract shape\cite{VT_shape} and design optimization of acoustic equipment.
%In addition, trained PINN is capable of making predictions on the scale of the millisecond scale \cite{PINN_Borrel_1D}, making it applicable to gaming, VR/AR applications, live performance, and other applications that require real-time performance. While standard finite difference or finite element methods can be used for the real-time applications, the storage requirements for calculations performed offline make them difficult \cite{PINN_Borrel_1D}\cite{PINN_Borrel_3D}. For these reasons, the development of PINNs that deal with acoustic resonance is highly significant.

In this study, we developed ResoNet, a PINN that analyzes acoustic resonance in the time domain based on the wave equation while effectively utilizing the function approximation capability of neural networks\cite{NN_UnivApprox} by training the neural network to minimize the loss function with respect to periodic solutions. The main contributions of this study are as follows: (i) this study devised a periodicity loss function that effectively utilizes the function approximation capability of neural networks when using the PINN for analyzing 1D resonances in the time domain; (ii) it entailed a detailed investigation of the applicability of PINN to 1D acoustic resonance analysis; and (iii) it examined the performance of inverse problem analysis on 1D acoustic resonance phenomena.

The remainder of this paper is organized as follows. Section \ref{sec:2} describes the one-dimensional wave equation with energy loss terms and the acoustic field setup analyzed in this study. Section \ref{sec:3} describes ResoNet, a PINN that analyzes resonances based on the wave equation described in Section \ref{sec:2}. Section \ref{sec:4} describes the forward and inverse analyses using ResoNet and its performance. Section \ref{sec:5} summarizes the study and discusses the application potential of PINN in acoustic resonance.

\section{\label{sec:2} Governing equations of acoustic resonance}
This section describes the wave equation and the acoustic field setup analyzed in this study.

\subsection{\label{subsec:2:1} One-dimensional wave equation with energy loss terms}
\begin{figure}
\includegraphics[width={8cm}]{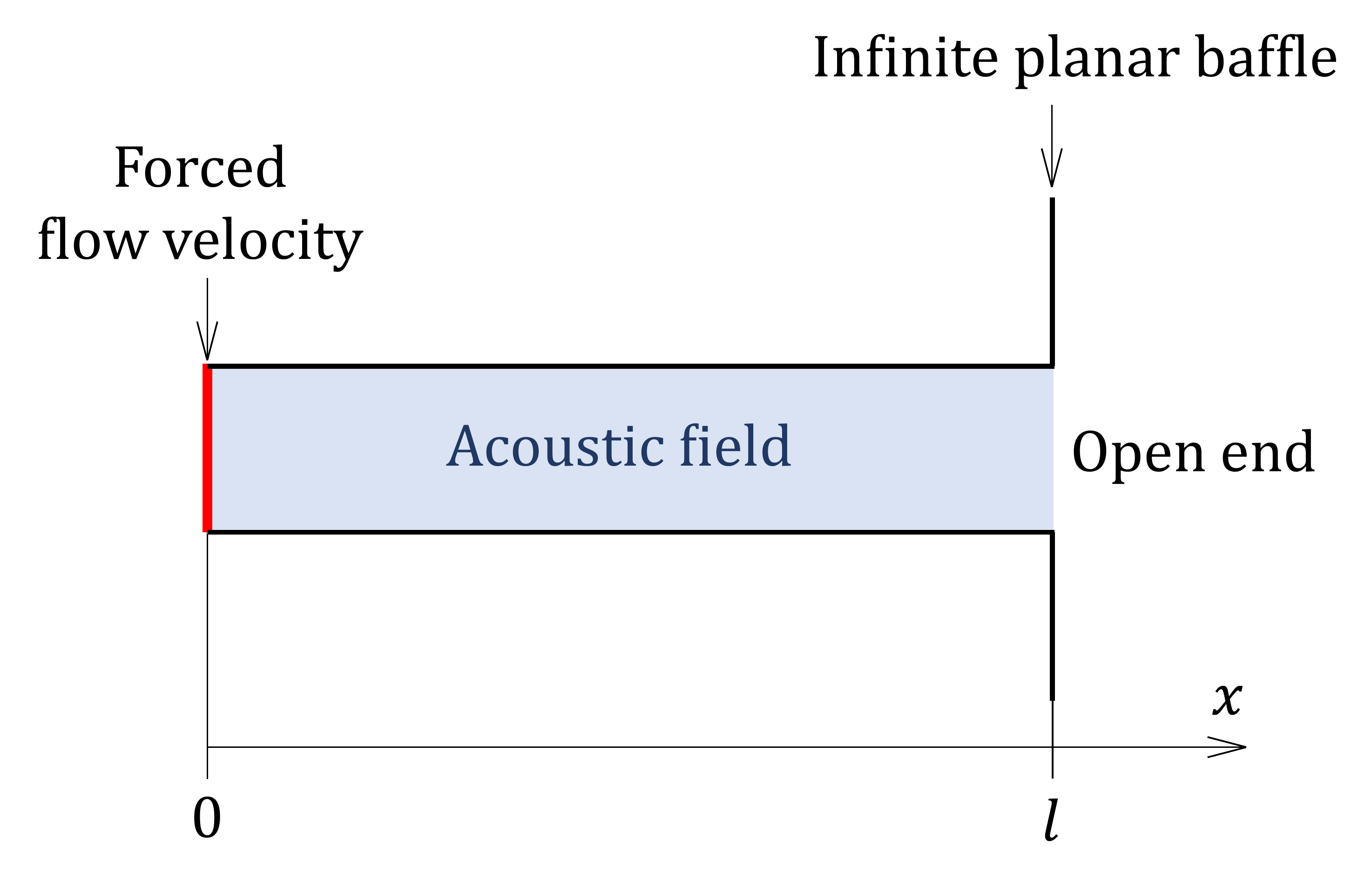}
\caption{\label{fig:FIG1}{(Color online) The 1D acoustic tube that is the subject of the analysis in this study. The flow boundary condition is given at the left end, and the right end is the open end. The radiation at the open end is modeled by the equivalent circuit shown in Fig. \ref{fig:FIG2}.}}
\end{figure}

We considered the propagation of plane sound waves in an acoustic tube of length $l$ and circular cross-sectional area of $A(x)$, as shown in Fig. \ref{fig:FIG1}, with $x$ as the axial direction. Let the sound pressure in the acoustic tube be $p$ and air volume velocity be $u$. Assuming $p=Pe^{j\omega t}$ and $u=Ue^{j\omega t}$, the telegrapher's equations for the acoustic tube considering energy loss are as follows\cite{2_TeleEq}.
\linenomath
\begin{align}
\dfrac{dU}{dx} &= -\left( G+j\omega \dfrac{A}{K}\right) P, \label{WaveEq_Freq1}\\
\dfrac{dP}{dx} &= -\left( R+j\omega \dfrac{\rho}{A}\right) U, \label{WaveEq_Freq2}
\end{align}
where $G$ is the coefficient of energy loss owing to thermal conduction at the tube wall; $R$ is the coefficient of energy loss owing to viscous friction at the tube wall; $j$ is the imaginary unit; $\omega$ is the angular velocity; $K$ is the bulk modulus; and $\rho$ is the air density. Equations (\ref{WaveEq_Freq1}) and (\ref{WaveEq_Freq2}) can be expressed in the time domain, as follows.
\begin{align}
\dfrac{\partial u}{\partial x} &= -Gp - \dfrac{A}{K}\dfrac{\partial p}{\partial t}, \label{WaveEq_Time1}\\
\dfrac{\partial p}{\partial x} &= -Ru - \dfrac{\rho}{A}\dfrac{\partial u}{\partial t}. \label{WaveEq_Time2}
\end{align}
The velocity potential $\phi$ is defined as:
\begin{align}
u &= -A\dfrac{\partial \phi}{\partial x} \label{VelPot1},\\
p &= RA\phi + \rho\dfrac{\partial \phi}{\partial t} \label{VelPot2}.
\end{align}
Note that $u$ is volume velocity. From equations (\ref{WaveEq_Time1})-(\ref{VelPot2}), we obtain the following wave equation with energy loss terms, as follows.
\begin{equation}
\dfrac{\partial ^{2}\phi }{\partial x^{2}}+\dfrac{1}{A}\dfrac{\partial A}{\partial x}\dfrac{\partial \phi }{\partial x}=GR\phi +\left( \dfrac{G\rho }{A}+\dfrac{RA}{K}\right) \dfrac{\partial \phi }{\partial t}+\dfrac{\rho }{K}\dfrac{\partial ^{2}\phi }{\partial t^{2}}.
\label{WaveEq}
\end{equation}
A detailed derivation of Eq. (\ref{WaveEq}) is provided in Appendix \ref{Ap_WaveEq}.

Theoretical solutions for $R$ and $G$ have been proposed, as follows, under the assumptions that the wall surface is rigid and thermal conductivity is infinite\cite{2_TeleEq}.
\begin{align}
R &= \dfrac{S}{A^2}\sqrt{\dfrac{\omega_c \rho \mu}{2}}, \label{TeleEq_R}\\
G &= S \dfrac{\eta -1}{\rho c^2}\sqrt{\dfrac{\lambda \omega_c}{2 c_p \rho}}, \label{TeleEq_G}
\end{align}
where $S$ is the circumference of the acoustic tube; $\mu$, $\eta$, $c$, $\lambda$, and $c_p$ are the viscosity coefficient, heat-capacity ratio, speed of sound, thermal conductivity, and specific heat at constant pressure, respectively; and $\omega_c$ is the angular velocity used to calculate the energy loss term. As $R$ and $G$ vary with the physical properties of air and the condition of the walls of the acoustic tube, these parameters are generally unknown. We confirmed that the energy loss parameters of the acoustic tube should be measured experimentally\cite{2_Yokota_VT}. Additionally, as demonstrated in vocal tract analysis, we have experimentally confirmed that the energy loss parameters can be approximated as frequency-independent constants in a limited frequency range\cite{2_Yokota_VT}. Therefore, in this study, $\omega_c$ was set as constant, and consequently, $R$ and $G$ were assumed to be frequency-independent constants.

\subsection{\label{subsec:2:2} Acoustic field and boundary conditions}
The acoustic field analyzed in this study is shown in light blue in Fig. \ref{fig:FIG1}. The acoustic tube was straight, with a forced-flow boundary condition at $x=0$. It had an open end with an infinite planar baffle at $x=l$, and the boundary condition was given by the equivalent circuit described in Section \ref{subsec:2:3}.

The aforementioned boundary conditions are fundamental in human speech production\cite{Time_2mass1,Time_2mass2} and in the acoustic analysis of brass instruments\cite{2_BrassInst}. In this study, we used these boundary conditions to investigate the performance of resonance analysis with PINN.

\subsection{\label{subsec:2:3} Modeling of radiation}
\begin{figure}
\includegraphics[width=\reprintcolumnwidth]{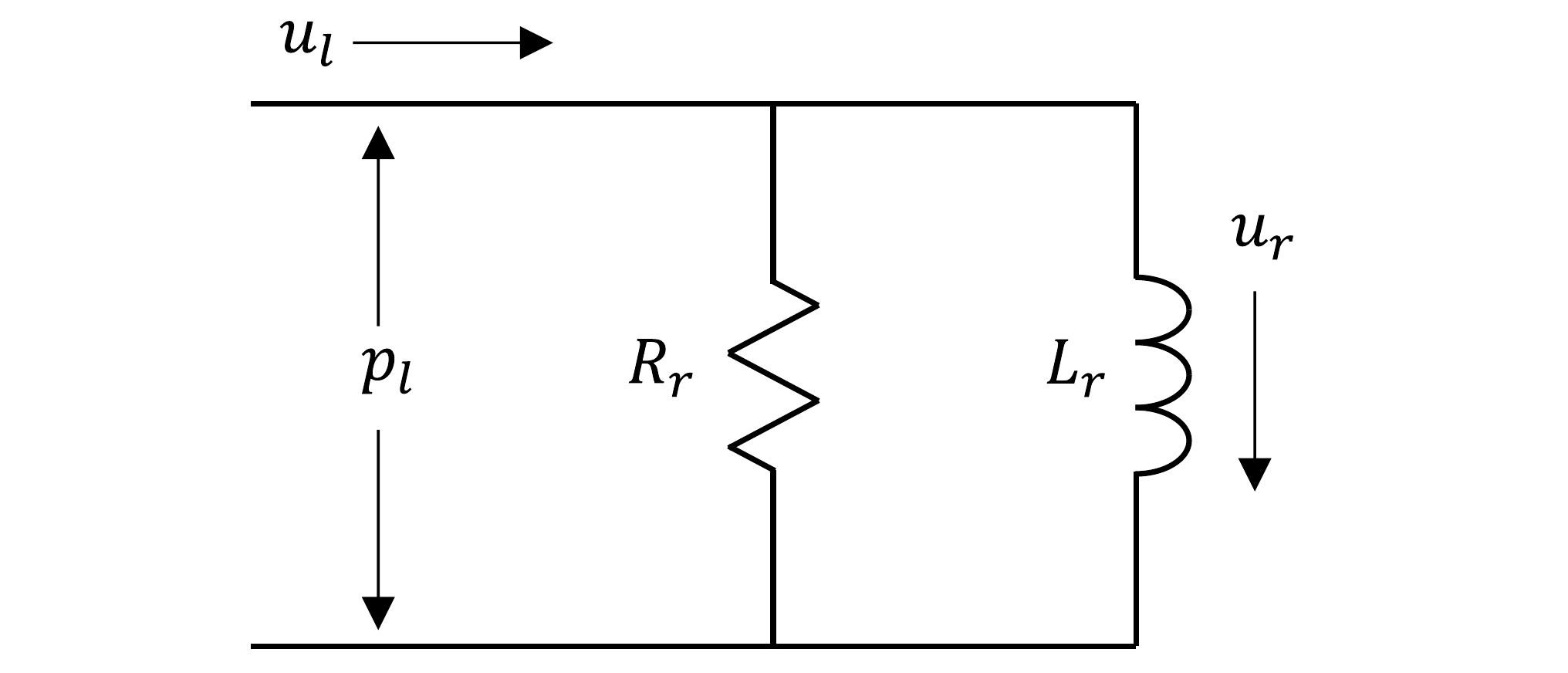}
\caption{\label{fig:FIG2}{Equivalent circuit of acoustic radiation. $u_l$ is the volume velocity at $x=l$. The volume velocity corresponds to the current, and the pressure corresponds to the voltage.}}
\end{figure}

This section describes the modeling of the radiation at $x=l$ in Fig. \ref{fig:FIG1}. Assuming that the particle velocity at the open end is uniform, air at the open end can be regarded as a planar sound source. In response to the acoustic radiation, the plane receives sound pressure $p_l$ from outside the acoustic tube. If an infinite planar baffle surrounds the open end, the relationship between volume velocity $u_l$ at the open end and $p_l$ can be approximated using the equivalent circuit shown in Fig. \ref{fig:FIG2}\cite{Time_2mass1}. In Fig. \ref{fig:FIG2}, the volume velocity $u_l$ corresponds to the current and the sound pressure $p_l$ to the voltage. The equations connecting $u_l$ and $p_l$ are as follows.
\begin{align}
\left( u_l - u_r \right) R_r &= L_r\dfrac{du_r}{dt}, \label{Radiation_EOM}\\
p_l &= \left( u_l - u_r \right) R_r, \label{Radiation_Pressure}
\end{align}
where $R_r$ is the circuit resistance; $L_r$ is the circuit reactance; and $u_r$ is the current (corresponding to the volume velocity) flowing through the coil side. Here, $R_r$ and $L_r$ are expressed as follows.
\begin{align}
R_r &= \dfrac{128 \rho c}{9 \pi^2 A_l},\\
L_r &= \dfrac{8 \rho}{3 \pi \sqrt{\pi A_l}},
\end{align}
where $A_l$ denotes the cross-sectional area for $x=l$.

\section{\label{sec:3} Proposed method}
This section describes the structure of ResoNet, a PINN for analyzing the acoustic resonance, and a training method for the neural network.

\subsection{\label{subsec:3:1} Overview of ResoNet}
\begin{figure*}
\includegraphics[width={17cm}]{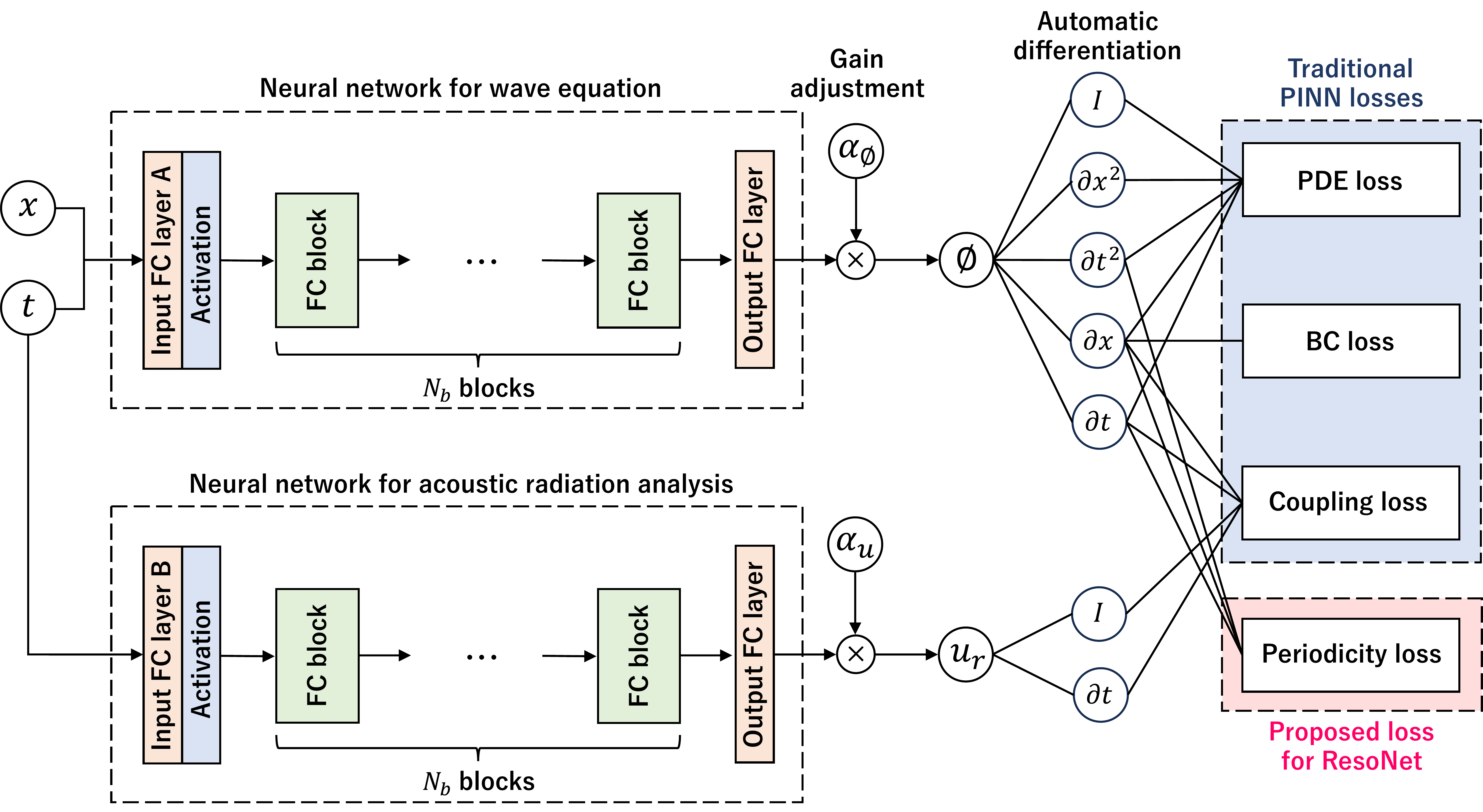}
\caption{\label{fig:FIG3}{(Color online) Structure of ResoNet. The upper neural network in the figure outputs the velocity potential $\phi$, and the lower neural network outputs the volume velocity $u_r$ with respect to the radiation. $\phi$ and $u_r$ are input to coupling loss, and coupled analysis is performed. Periodicity Loss is the loss function proposed in this study, which forces $u$, $p$, and $\phi$ to be the same at $t=0$ and $t=T$ ($T$: period).}}
\end{figure*}

Fig. \ref{fig:FIG3} shows the proposed neural network structure for the resonance analysis in acoustic tubes, referred to as ResoNet. As shown in Fig. \ref{fig:FIG3}, ResoNet has two blocks of neural networks.

The first is a network that calculates the solutions to the wave equation in Eq. (\ref{WaveEq}), taking $x_i$ and $t_i$ as inputs (where $i$ is the sample number) to predict the velocity potential $\hat{\phi}_i$. In this study, this input--output relationship is expressed using the following equation:
\begin{equation}
\hat{\phi}_i = F_{w} \left( x_i, t_i; \Theta_{w} \right),
\label{Func_WaveEq}
\end{equation}
where $F_w$ is the operator of the neural network for the wave equation, and $\Theta_w$ is the set of trainable parameters of the neural network.

The other is a network for calculating the acoustic radiation, utilizing $t_i$ as the input for predicting $\hat{u}_{ri}$ by calculating the solution of the equivalent circuit in Fig. \ref{fig:FIG2}. The input-output relationship is expressed as:
\begin{equation}
\hat{u}_{ri} = F_r \left( t_i; \Theta_r \right),
\label{Func_Rad}
\end{equation}
where $F_r$ is the neural network operator for acoustic radiation analysis, and $\Theta_r$ is the set of trainable neural network parameters.

The reason for using two neural networks instead of one with two outputs is to modify the neural network of the external system (in this paper, acoustic radiation system). The vocal fold vibration system in speech production analysis and the spatial acoustic system in instrumental acoustics can be coupled with tube acoustics. Each PINN simulating the respective physical system has its suitable structure and scale (e.g., choice of activation function). Therefore, in this study, we separated the wave equation system and the acoustic radiation system into separate neural networks to facilitate modification of the neural network for the external system.

In this study, all activation functions used in the neural network shown in Fig. \ref{fig:FIG3} are functions expressed as
\begin{equation}
f(a) = a+\sin ^2 a,
\label{Activation}
\end{equation}
where $a$ denotes the input to the layer. This activation function, Snake, has been reported to be robust to periodic inputs\cite{3_Snake}. Snake was selected as the activation function because of the better learning performance compared to the more traditional tanh and sin in preliminary simulations. Considerations regarding the choice of activation function are discussed in Appendix \ref{Ap_Activation}.

We performed automatic differentiation on $\hat{\phi}_i$ and $\hat{u}_{ri}$, and the various loss functions shown on the right side of Fig. \ref{fig:FIG3} are defined. The traditional PINN uses the PDE and boundary condition (BC) loss functions. The PDE loss introduces the constraints of Eq. (\ref{WaveEq}), and the BC loss introduces constraints owing to the boundary conditions into the neural network. Coupling loss is a type of PDE loss in traditional PINN, which introduces constraints owing to the coupling of the wave equation and the external system (acoustic radiation in this study) into the neural network, which is a type of PDE loss for conventional PINNs. We used periodicity loss that introduced constraints owing to the resonance periodicity. The following sections describe each item in detail.

\subsection{\label{subsec:3:2} Neural network blocks}
As shown in Fig. \ref{fig:FIG3}, ResoNet has two neural network blocks: one calculates the solution of the wave equation in Eq. (\ref{WaveEq}), whereas the other calculates the solution for the acoustic radiation circuit in Fig. \ref{fig:FIG2}.

\subsubsection{\label{subsec:3:2:1} Neural network for wave equation}
This network uses $x_i$ and $t_i$ as inputs and predicts the velocity potential $\hat{\phi}_i$ for the wave equation. Initially, two-channel data ($x_i$, $t_i$) are fed to ``Input FC layer A," which is a fully connected layer\cite{3_Introduction_NN} that outputs $N_f$ channel data as Fig. \ref{fig:FIG3}shows. An activation layer is present immediately thereafter, and it is the snake described in Section \ref{subsec:3:1}.

\begin{figure}
\includegraphics[width=\reprintcolumnwidth]{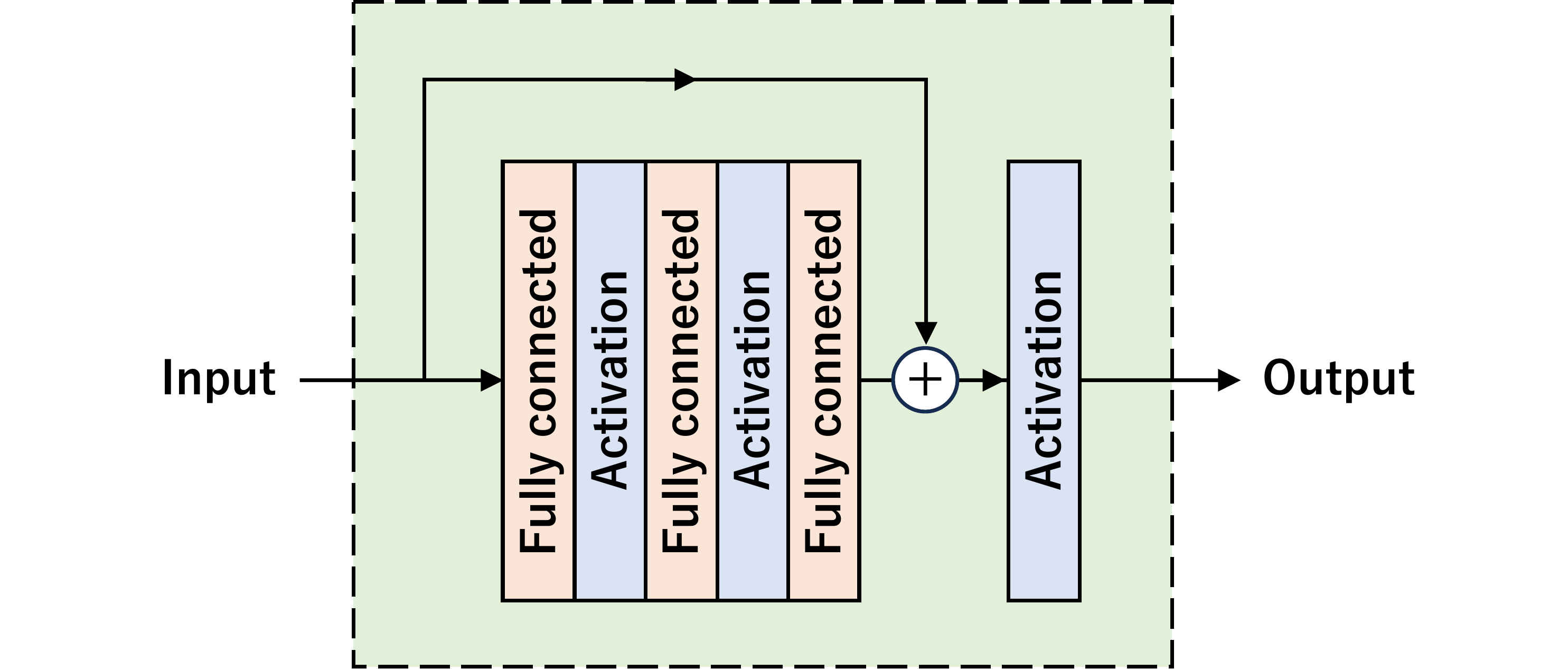}
\caption{\label{fig:FIG4}{(Color online) Details of FC block. There are two paths in the block, one calculated by fully connected layers and activation functions and the other by skip paths. All activation functions are 'snake'.}}
\end{figure}

After the first activation layer, data are fed into the ``FC block.” All FC blocks in Fig. \ref{fig:FIG3} have the same structure, and its details are shown in Fig. \ref{fig:FIG4}. The fully connected layer in Fig. 4 has $N_f$ input channels and $N_f$ output channels, and the activation layer is the Snake, as shown in Eq. (\ref{Activation}). To circumvent the vanishing gradient problem\cite{3_Vanishing}, a residual connection\cite{3_ResNet} is applied before the last activation layer.

After $N_b$ FC blocks, we fed the data into the ``Output FC layer, " a fully connected layer with $N_f$ input channels and one output channel. The output value is multiplied by the gain adjustment parameter $\alpha_{\phi}$, and finally the neural network predicts the $\hat{\phi}_i$ value of the solution for the wave equation. 

\subsubsection{\label{subsec:3:2:2} Neural network for acoustic radiation analysis}
This network uses $t_i$ as the input and predicts the $\hat{u}_{ri}$ value of the equivalent circuit shown in Fig. \ref{fig:FIG2}. The only difference from the neural network for the wave equation (described in Section \ref{subsec:3:2:1}) is that the number of input channels is one ($t_i$ only), and the ``Input FC layer B" is a fully connected layer with one input channel and $N_f$ output channels. All the other structures are identical to those described in Section \ref{subsec:3:2:1}.

\subsection{\label{subsec:3:3} Loss functions}
The loss function of ResoNet is calculated as the sum of the following partial loss functions: traditional PINN losses (PDE loss, BC loss and Coupling loss), and periodicity loss.

\subsubsection{\label{subsec:3:3:1} Traditional PINN losses}
As regards PDE loss, the output $\hat{\phi}_i$ of the neural network is defined as:
\begin{equation}
\hat{\phi}_{i,E} := F_{w} \left( x_i, t_i; \Theta_{w} \right),
\quad x_i \in \left[ 0,l \right], \quad t_i \in \left[ 0,T \right],
\label{Phi_PDE}
\end{equation}
where $T$ is the simulation time ( one resonance period in this study). For $\hat{\phi}_{i,E}$ to follow Eq. (\ref{WaveEq}), the PDE loss is defined as:
\begin{equation}
\begin{split}
L_E &= \dfrac{1}{N_{E}} \sum_{i=1}^{N_{E}} \left\{ \dfrac{\partial ^{2}\hat{\phi}_{i,E} }{\partial x_i^{2}}+\dfrac{1}{A_i}\dfrac{\partial A_i}{\partial x_i}\dfrac{\partial \hat{\phi}_{i,E} }{\partial x_i} - G_iR_i\hat{\phi}_{i,E} \right. \\
&\left. \quad -\left( \dfrac{G_i\rho}{A_i}+\dfrac{R_iA_i}{K}\right) \dfrac{\partial \hat{\phi}_{i,E} }{\partial t_i}-\dfrac{\rho}{K}\dfrac{\partial ^{2}\hat{\phi}_{i,E} }{\partial t_i^{2}} \right\}^2,
\label{Loss_PDE}
\end{split}
\end{equation}
where $N_E$ is the number of collocation points for the PDE loss. The partial differential values of $\hat{\phi}_{i,E}$ are required to calculate Eqs. (\ref{Loss_PDE}) are obtained via automatic differentiation\cite{3_Vanishing} of the neural network.

Similarly, $\hat{\phi}_i$ must follow boundary conditions. As described in Section \ref{subsec:2:2}, the boundary condition in this study at $x=0$ is given by the forced flow velocity. For the BC loss, the output $\hat{\phi}_i$ of the neural network is defined as:
\begin{equation}
\hat{\phi}_{i,B} := F_{w} \left( x_0, t_i; \Theta_{w} \right),
\quad t_i \in \left[ 0,T \right],
\label{Phi_BC}
\end{equation}
where $x_0=0$. The loss function $L_{B}$ with respect to the boundary condition is defined as:
\begin{equation}
L_{B} = \dfrac{1}{N_B} \sum_{i=1}^{N_B} \left( \hat{u}_{i,B} - \bar{u}_{i,B} \right) ^2,
\label{Loss_BC}
\end{equation}
where $N_B$ is the number of collocation points for the BC loss and $\bar{u}_{i,B}$ is the volume flow velocity data given as the boundary condition. Based on Eq. (\ref{VelPot1}), $\hat{u}_{i,B}$ is calculated from $\hat{\phi}_{i,B}$ using the following equation.
\begin{equation}
\hat{u}_i = -A_i\dfrac{\partial \hat{\phi}_i}{\partial x_i}.
\label{Calc_u}
\end{equation}

\subsubsection{\label{subsec:3:3:2} Coupling loss (A type of PDE loss)}
ResoNet incorporates the coupling condition between the system described by the wave equation and the external system as the coupling loss. Coupling loss is a type of PDE loss in traditional PINN. In this study, the external system is an acoustic radiation system, as shown in Fig. \ref{fig:FIG2}, and the Eqs. (\ref{Radiation_EOM}) and (\ref{Radiation_Pressure}) describe coupling. In this section, we describe a method for calculating the coupling loss.

First, we define the output $\hat{\phi}_i$ of the neural network for the wave equation as:
\begin{equation}
\hat{\phi}_{i,C} := F_{w} \left( x_l, t_i; \Theta_{w} \right), \quad
\quad t_i \in \left[ 0,T \right],
\label{Out_coupling}
\end{equation}
where $x_l=l$. Based on Eq. (\ref{VelPot2}), the sound pressure is obtained from $\hat{\phi}_i$ as follows.
\begin{equation}
\hat{p}_i = R_iA_i\hat{\phi}_i + \rho\dfrac{\partial \hat{\phi}_i}{\partial t_i}. \label{Calc_p}
\end{equation}
Let $\hat{u}_{i,C}$ and $\hat{p}_{i,C}$ be the volume velocity and sound pressure at $x=l$, obtained by applying Eqs. (\ref{Calc_u}) and (\ref{Calc_p}) to $\hat{\phi}_{i,C}$, respectively. Additionally, as defined in Eq. (\ref{Func_Rad}), let $\hat{u}_{ri}$ be the output of the neural network for acoustic radiation for input $t_i$. Based on Eqs. (\ref{Radiation_EOM}) and (\ref{Radiation_Pressure}), the coupling loss $L_C$ is defined as:
\begin{equation}
\begin{split}
L_C &= \dfrac{\lambda_l}{N_C} \sum_{i=1}^{N_C}
\left\{ \left( \hat{u}_{ri} - \hat{u}_{i,C} \right) R_r - L_r\dfrac{d\hat{u}_{ri}}{dt_i} \right\}^2 \\
&\quad + \dfrac{\lambda_r}{N_C} \sum_{i=1}^{N_C} \left\{ \hat{p}_{i,C} - \left( \hat{u}_{ri} - \hat{u}_{i,C} \right) R_r \right\}^2,
\end{split}
\end{equation}
where $N_C$ is the number of collocation points of the coupling loss; and $\lambda_l$ and $\lambda_r$ are the weight parameters of each term.

\subsubsection{\label{subsec:3:3:3} Periodicity loss}
This section explains the core idea of ResoNet, the loss function with respect to periodicity. When performing time-domain resonance analysis, the transient state must be included until a steady state is reached. As mentioned in Section \ref{sec:1}, because sound waves have complex dynamics, a large-scale neural network is required to perform long simulations, which poses challenges in the computational cost and neural network convergence.

\begin{figure}
\includegraphics[width=\reprintcolumnwidth]{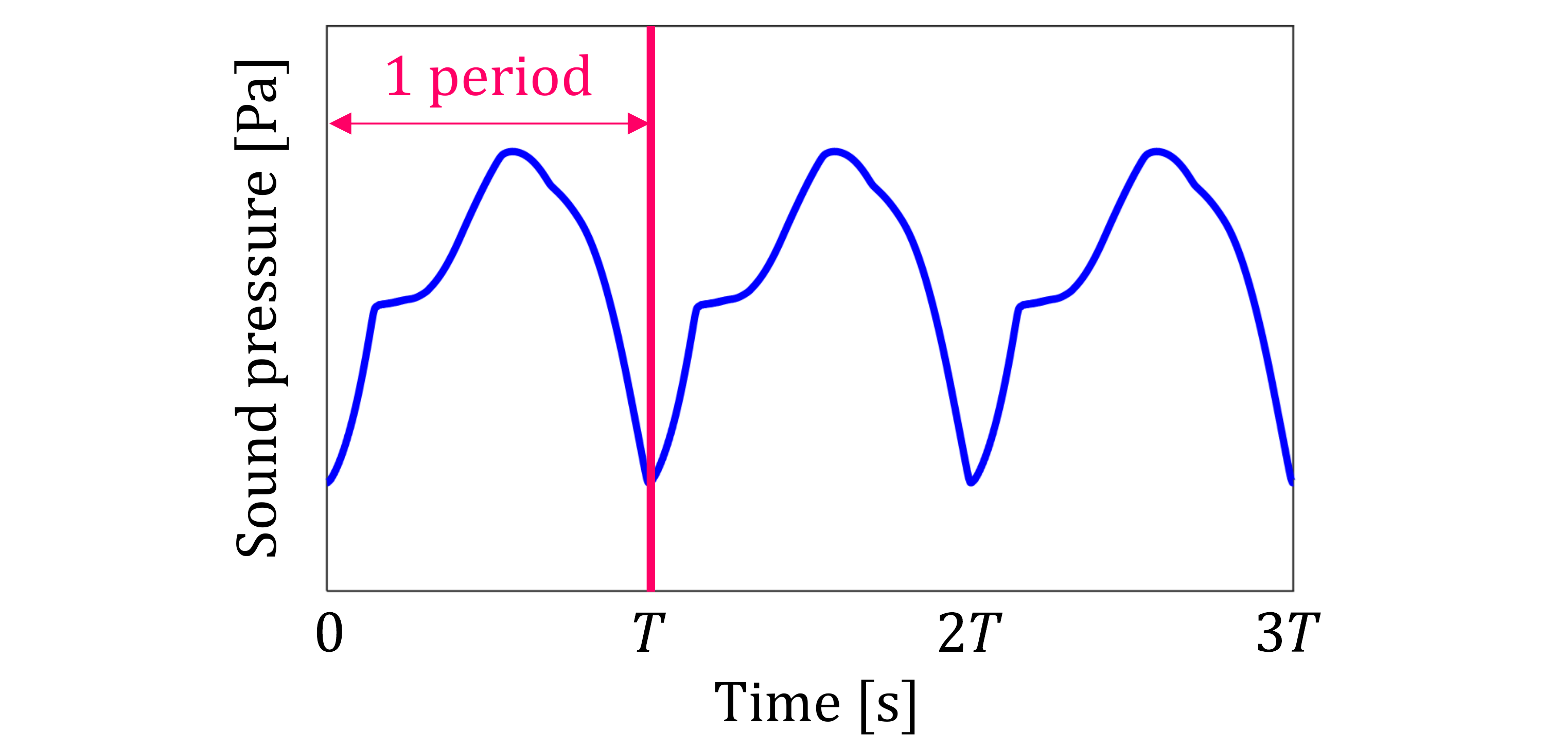}
\caption{\label{fig:FIG5}{(Color online) Sound pressure waveform at steady state of resonance. In the steady state of the resonance, the waveform changes little in each period.}}
\end{figure}

In the acoustic resonance analysis, the object of interest is one period in the steady state, as shown in Fig. \ref{fig:FIG5}. Therefore, in ResoNet, only one period was analyzed, and the function approximation capability of the neural network was effectively utilized. For this purpose, we formulated a ``periodicity loss” function, which forces the output of the neural network to have the same value as $\hat{\phi}_i$ at $t=0$ and $t=T$ in the steady state. The following describes the procedure for calculating periodicity loss.

First, the output $\hat{\phi}_i$ of the neural network for the wave equation is defined using the following two equations.
\begin{equation}
\begin{array}{c}
\hat{\phi}_{i,P1} := F_w \left( x_i, t_0; \Theta_w \right) \\
\hat{\phi}_{i,P2} := F_w \left( x_i, t_T; \Theta_w \right)
\end{array}
, \quad x_i \in \left[ 0,l \right],
\label{Phi_P}
\end{equation}
where $t_0=0$ and $t_T=T$ ($T$: period); and $\phi_{i,P1}$ and $\phi_{i,P2}$ are the values of the output $\hat{\phi}_i$ for times $t=0$ and $t=T$ at the same position $x_i$, respectively.

The volume velocity is obtained from $\hat{\phi}_i$ using Eq. (\ref{Calc_u}), and the sound pressure is obtained from $\hat{\phi}_i$ using Eq. (\ref{Calc_p}). Let $\hat{u}_{i,P1}$ and $\hat{p}_{i,P1}$ be the volume velocity and sound pressure obtained from $\hat{\phi}_{i,P1}$, and let $\hat{u}_{i,P2}$ and $\hat{p}_{i,P2}$ be those obtained from $\hat{\phi}_{i,P2}$, respectively.
In the steady state, if the input $x_i$ (position) is the same, the continuity condition must hold between $\hat{p}_{i,P1}$ and $\hat{p}_{i,P2}$ and between $\hat{u}_{i,P1}$ and $\hat{u}_{i,P2}$. Therefore, the following loss functions are defined to force the neural network to enforce the conditions $\hat{p}_{i,P1}=\hat{p}_{i,P2}$ and $\hat{u}_{i,P1}=\hat{u}_{i,P2}$.
\begin{align}
L_u &= \dfrac{1}{N_P} \sum_{i=1}^{N_P} \left( \hat{u}_{i,P1} - \hat{u}_{i,P2} \right) ^2 \label{Loss_P_u} \\
L_p &= \dfrac{1}{N_P} \sum_{i=1}^{N_P} \left( \hat{p}_{i,P1} - \hat{p}_{i,P2} \right) ^2 \label{Loss_P_p},
\end{align}
where $N_P$ is the number of collocation points for the periodicity loss.
Additionally, we define the following loss function to enforce the continuous condition of the time derivative.
\begin{equation}
L_t = \dfrac{1}{N_P} \sum_{i=1}^{N_{P}} \left(
\dfrac{\partial ^{2}\hat{\phi}_{i,P1} }{\partial t_0^{2}} - \dfrac{\partial ^{2}\hat{\phi}_{i,P2} }{\partial t_T^{2}} \right) ^2.
\label{Loss_P_dt2}
\end{equation}

From Eqs. (\ref{Loss_P_u})–(\ref{Loss_P_dt2}), the periodicity loss $L_P$ proposed in this study is defined as
\begin{equation}
L_P = \lambda_u L_u + \lambda_p L_p + \lambda_t L_t,
\label{Loss_P}
\end{equation}
where $\lambda_u$, $\lambda_p$ and $\lambda_t$ are the weight parameters for each term.

If the periodicity loss is not included, the proposed method can analyze non-periodic phenomena in a 1D acoustic tube with radiation at the open end.

\subsubsection{\label{subsec:3:3:4} Loss function for the whole network}
The loss function $L_{all}$ for the entire network was calculated as the sum of the traditional PINN losses, periodic loss, and coupling loss, as follows.
\begin{equation}
L_{all} = \lambda_E L_E + \lambda_B L_B + \lambda_P L_P + \lambda_C L_C,
\label{Loss_all}
\end{equation}
where $\lambda_E$, $\lambda_B$, $\lambda_P$, and $\lambda_C$ denote the weight parameters of the respective loss functions. Finally, we formulated the optimization problem of ResoNet as follows:
\begin{equation}
\min_{\Theta_w,\Theta_r} L_{all}(\Theta_w,\Theta_r).
\end{equation}
By minimizing the loss function $L_{all}$, we optimized the trainable parameters $\Theta_w$ and $\Theta_r$ of the neural network. For this purpose, we used the Adam optimizer\cite{3_Adam} to determine the optimal values of $\Theta_w$ and $\Theta_r$ through iterative calculations.

\subsection{\label{subsec:3:4} Implementation}
We implemented ResoNet using the Deep Learning Toolbox in MATLAB (MathWorks, USA) and used the ``dlfeval" function to code a custom training loop. We used ``sobolset" function from the Statistics and Machine Learning Toolbox to create datasets for $x_i$ and $t_i$. The neural network was trained via GPU-assisted computation using the Parallel Computing Toolbox.

We performed the neural network training and prediction on a computer with a Core i9-13900KS CPU (Intel, USA) and GeForce RTX 4090 GPU (NVIDIA, USA) with 128 GB of main memory and 24 GB of video memory.

\section{\label{sec:4} Validation of proposed method}
The performance of the proposed method was validated through forward and inverse analysis of the acoustic resonance using ResoNet.

\subsection{\label{subsec:4:1} Forward analysis}
\subsubsection{\label{subsec:4:1:1} Analysis conditions for forward analysis}
The effectiveness of the proposed method was assessed through a forward analysis of the acoustic tube, as shown in Fig. \ref{fig:FIG1} using the boundary conditions described in Section \ref{subsec:2:2}. The length of the acoustic tube, $l$, was set to 1 m, and the diameter to 10 mm.

\begin{figure}
\includegraphics[width={7cm}]{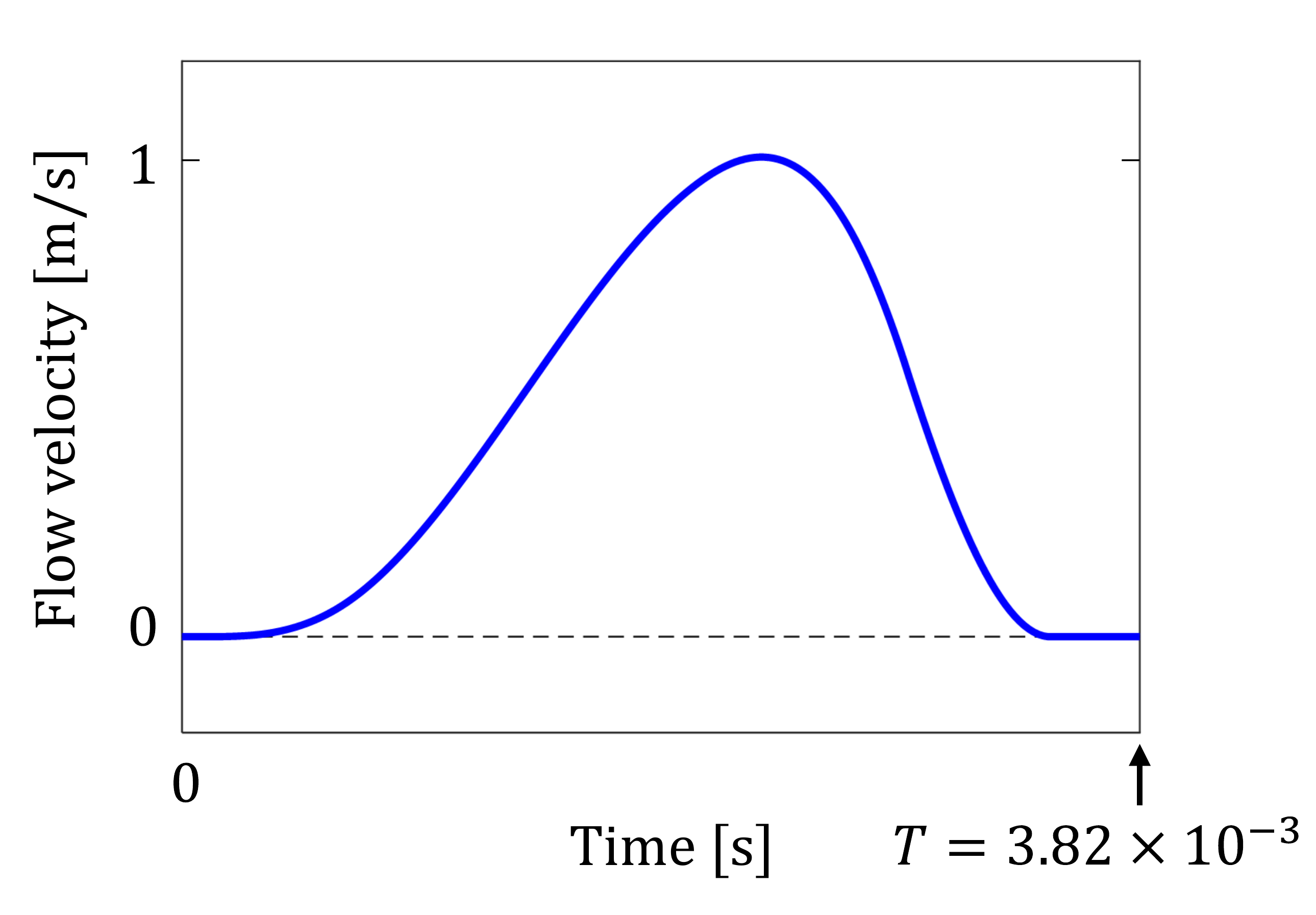}
\caption{\label{fig:FIG6}{(Color online) Forced flow velocity waveform (smoothed Rosenberg waveform). This waveform gives the boundary condition of the acoustic tube at $x=0$.}}
\end{figure}

The forced-flow velocity waveform, given as a boundary condition, is a smoothed Rosenberg wave\cite{3_Rwave} as shown in Fig. \ref{fig:FIG6}. A moving average filter was applied to the original Rosenberg wave to smooth the waveform, similar to an R++ wave\cite{3_Rwave_avg}. The fundamental frequency $F_0$ of the forced flow waveform was 261.6 Hz (C4 on the musical scale). Therefore, $T=3.82 \times 10^{-3}$ s.

The physical properties of air used in the analysis are listed in Table \ref{Table1}. Ishizaka et al. calculated $R$ by substituting a constant for $\omega_c$ in Eqs for the energy loss coefficients. (\ref{TeleEq_R})\cite{Time_2mass1}. In this study, we calculated $R$ and $G$ by substituting $\omega_c =1643.7$ rad/s (261.6 Hz, C4 in the musical scale) in Eqs. (\ref{TeleEq_R}) and (\ref{TeleEq_G}), respectively; thus, we obtained $R=6.99\times 10^5$ $\mathrm{m^2/(Pa\cdot s)}$ and $G=3.65\times 10^{-7}$ $\mathrm{Pa\cdot s/m^4}$.
\begin{table}[ht]
\caption{\label{Table1}Physical properties of air. The physical properties shown here are used in all forward and inverse analyses in this study.}
\centering
\begin{ruledtabular}
\begin{tabular}{cc}
Parameter&Value\\
\hline
Air density $\rho$&1.20 $\mathrm{kg/m^3}$\\
Bulk modulus $K$&$1.39\times10^5$ $\mathrm{Pa}$\\
Speed of sound $c$&340 $\mathrm{m/s}$\\
Viscosity coefficient $\mu$&19.0$\times10^{-6}$ $\mathrm{Pa\cdot s}$\\
Heat capacity ratio $\eta$&1.40\\
Thermal conductivity $\lambda$&2.41$\times10^{-2}$ $\mathrm{W/(m\cdot K)}$\\
Specific heat for const. pressure $c_p$&1.01 $\mathrm{kJ/(kg\cdot K)}$\\
\end{tabular}
\end{ruledtabular}
\end{table}

In the forward analysis, we set the number of nodes in the neural network $N_f$ to 200 and the number of FC blocks $N_b$ to five based on preliminary simulations. Further, we created a dataset ($x_i$, $t_i$) for each loss function, as follows. In Eq. (\ref{Phi_PDE}) for the PDE loss calculation, we used quasi-random numbers generated by the ``sobolset" function in MATLAB for $x_i$ and $t_i$, and the number of collocation points $N_E$ was 5000. Next, for $t_i$ in Eq. (\ref{Phi_BC}), and Eq. (\ref{Out_coupling}) used in the calculation of the BC and coupling losses, the range $[0,T]$ was divided into 1000 equal parts to create a dataset of $t_i$; thus, $N_B$ and $N_C$ were 1000. Finally, for $x_i$ in Eqs. (\ref{Phi_P}) used in calculating the periodicity loss, the range $[0,l]$ was divided into 1000 equal parts to create a dataset of $x_i$; thus, $N_P$ was 1000. The validity of the number of these collocation points is explained in Appendix \ref{Ap_Collocation}. As in Rasht-Behesht et al.\cite{PINN_Seismic2}, the value ranges of $x_i$ and $t_i$ were normalized to $[-1,1]$ when inputting them into the neural network. The gain adjustment parameters were set to $\alpha_{\phi}=0.002$ and $\alpha_u=3.4 \times 10^{-5}$.

To compare the performance of the forward analysis, we used the finite difference method. The step size of the finite difference method was $\Delta x=10^{-3}$ m, $\Delta t=0.5 \times 10^{-6}$ s, and we used the CTCS (Centered-Time Centered-Space) scheme in the calculations. Appendix \ref{Ap_FDM} describes the validity of the step size.

\subsubsection{\label{subsec:4:1:2} Hyperparameters of Neural Network}
This section describes the PINN weight parameters and the learning rate of the Adam optimizer.

Setting appropriate weight parameters is critical for PINN learning. In this study, we determined the weight parameters by using a combination of two approaches: a normalization approach based on known boundary conditions and an empirical approach. First, for the known boundary condition at $x=0$, the maximum particle velocity is 1 m/s, as shown in Fig. \ref{fig:FIG6}; therefore, the weight $\lambda_B$ with respect to the boundary condition is normalized as follows
\begin{equation}
\lambda_B = \frac{\lambda'_B}{A_0^2}
\label{Lambda_B}
\end{equation}
where $A_0$ is the cross-sectional area at $x=0$, and we set $\lambda'_B=1$. With this $\lambda'_B$ as a reference, the weight parameters $\lambda_u$, $\lambda_l$, and $\lambda_t$, which are related to the volume velocity, were normalized as follows.
\begin{align}
\lambda_u &= \dfrac{\lambda'_u}{A_0^2} \label{lambda_u} \\
\lambda_l &= \dfrac{\lambda'_l}{\left( R_r A_0 \right)^2} \label{lambda_l} \\
\lambda_r &= \dfrac{\lambda'_r}{\left( R_r A_0 \right)^2} \label{lambda_r},
\end{align}
Since the order of the particle velocity is nearly constant in the tube, $\lambda'_B$, $\lambda'_u$, $\lambda'_l$ and $\lambda'_r$ should have close orders. Finally, we empirically set $\lambda'_u=1$, $\lambda'_l=1$, and $\lambda'_r=50$. The remaining $\lambda_E$, $\lambda_p$, and $\lambda_t$ are weights with respect to pressure $p$ and velocity potential $\phi$. Therefore, we cannot infer the order from the boundary conditions. Hence, we empirically set them to $\lambda_E=0.58$, $\lambda_p=8.7 \times 10^{-6}$, and $\lambda_t=1.3 \times 10^{-12}$. The final list of weight parameters used is shown in Table \ref{Table2}.
\begin{table}[ht]
\caption{\label{Table2}Weight parameters for forward analysis. We determined these parameters by a combination of normalization and empirical approaches described in Section \ref{subsec:4:1:2}.}
\centering
\begin{ruledtabular}
\begin{tabular}{cccc}
Parameter&Value&Parameter&Value\\
\hline
$\lambda_E$&$0.58$&$\lambda_r$&$1.4 \times 10^{-4}$\\
$\lambda_B$&$1.6 \times 10^8$&$\lambda_u$&$1.6 \times 10^8$\\
$\lambda_P$&$1.0$&$\lambda_p$&$8.7 \times 10^{-6}$\\
$\lambda_C$&$1.0$&$\lambda_t$&$1.3 \times 10^{-12}$\\
$\lambda_l$&$2.9 \times 10^{-6}$\\
\end{tabular}
\end{ruledtabular}
\end{table}

We trained all collocation points as a single batch, and the learning rate of the Adam optimizer $\lambda_{\rm{adam}}$ was calculated by the following equation.
\begin{equation}
    \lambda_{\rm{adam}} = \frac{\lambda_{\rm{init}}}{1+ \beta i_e}
\end{equation}
where $\lambda_{\rm{init}}$ is the initial value of the learning rate, $\beta$ is the decay coefficient, and $i_e$ is the number of epochs. In this study, we set $\lambda_{\rm{init}}=0.001$ and $\beta=0.007$ empirically.

\subsubsection{\label{subsec:4:1:3} Results of forward analysis}
\begin{figure}[h]
\baselineskip=12pt
\figcolumn{
\fig{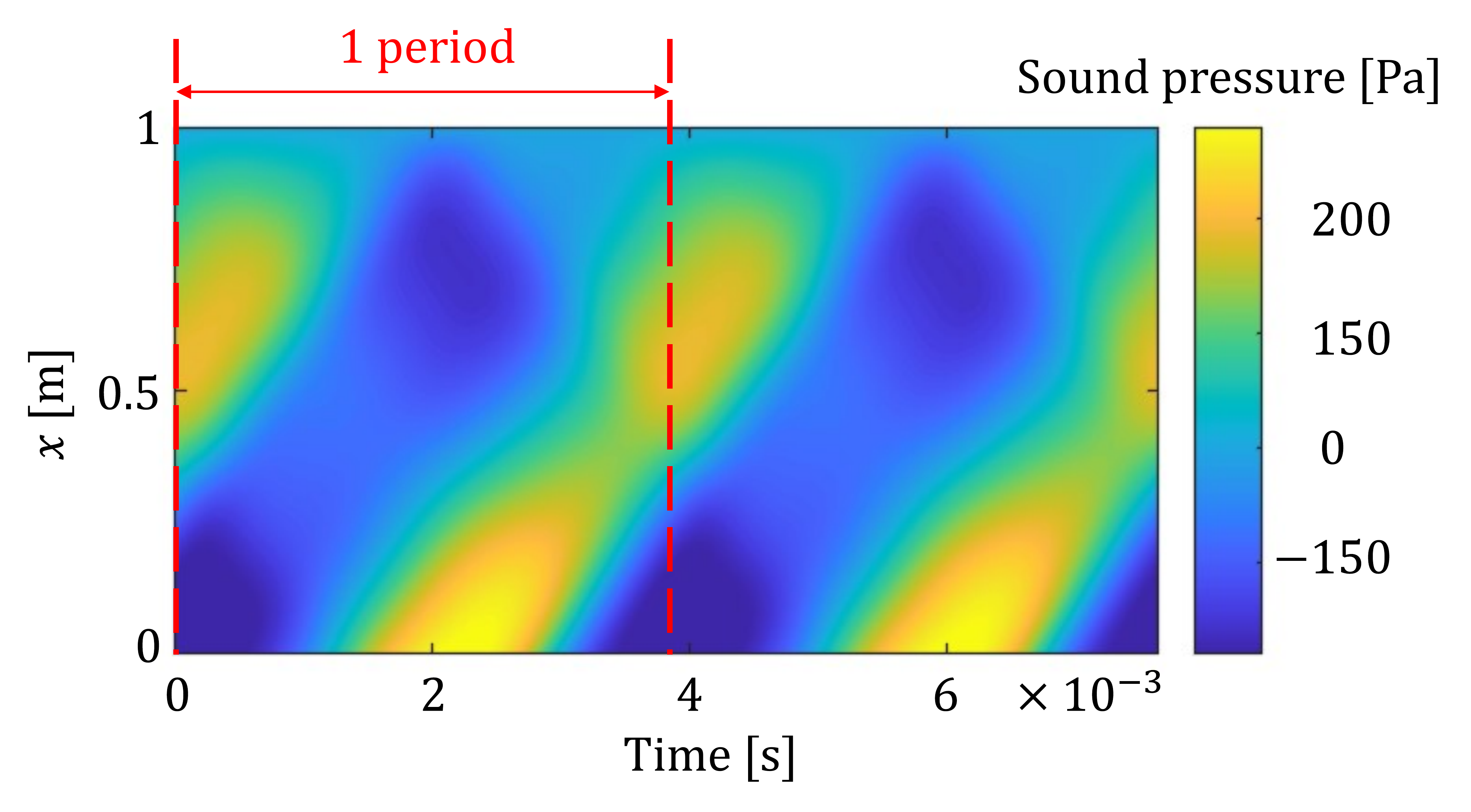}{8cm}{(a) Result of ResoNet.}
\fig{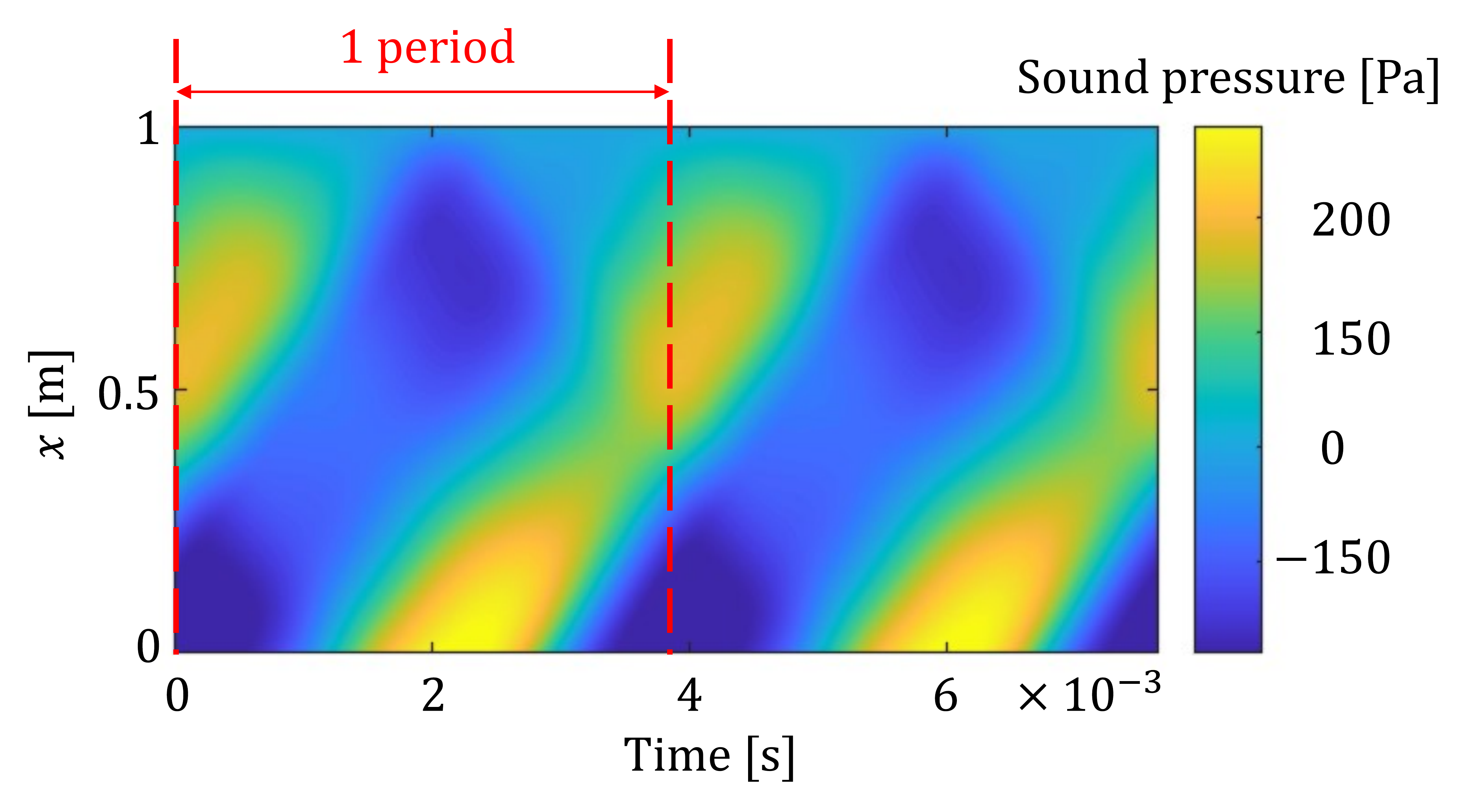}{8cm}{(b) Result of FDM.}
}
\caption{\label{fig:FIG7}(Color online) Analyzed sound pressure. The results of the ResoNet forward analysis are in close agreement with the Finite Difference Method (FDM) analysis results.}
\end{figure}

\begin{figure}
\includegraphics[width={8cm}]{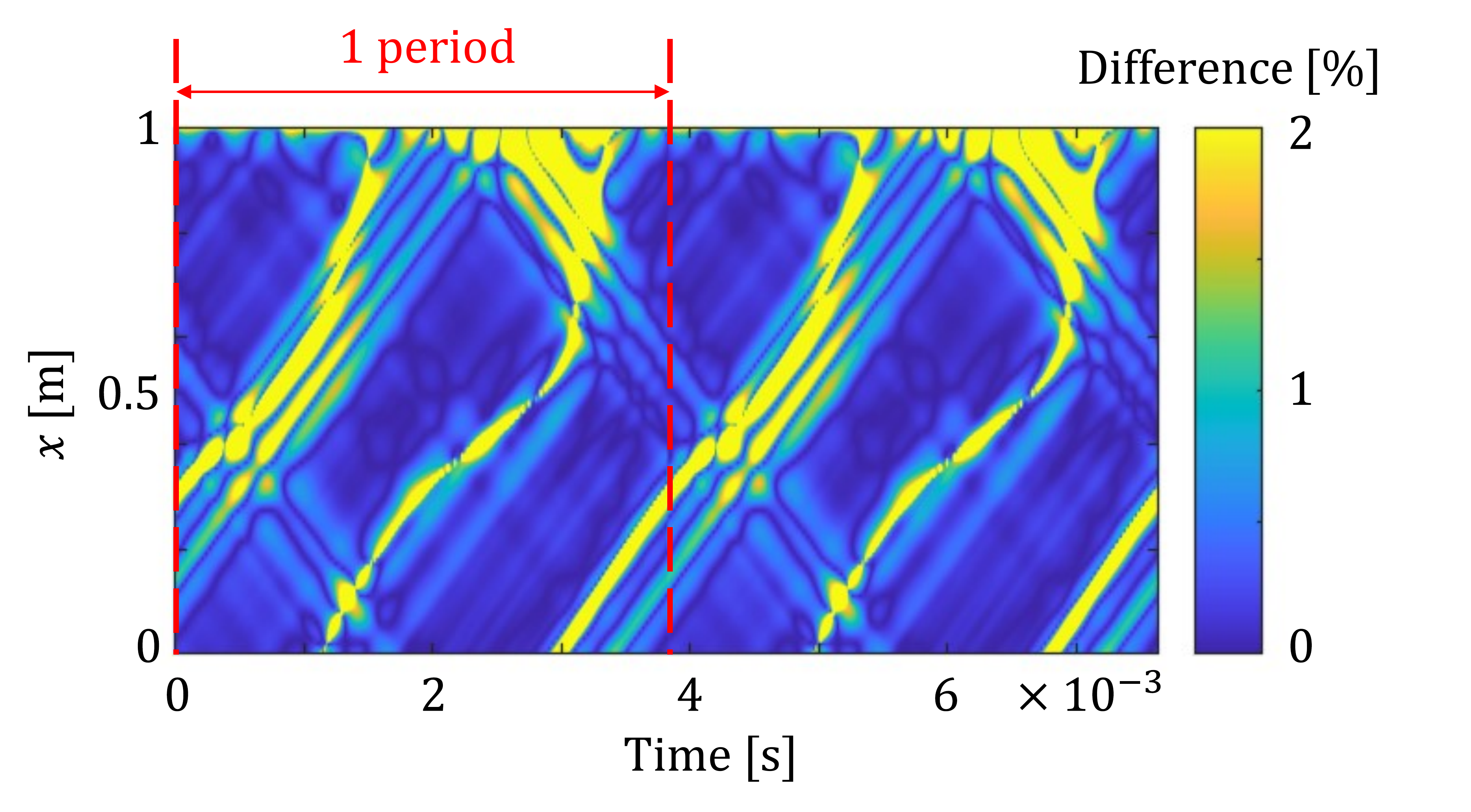}
\caption{\label{fig:FIG8}{(Color online) Difference of the ResoNet from the FDM. In most regions, the difference is less than 1\%, but there are some regions where the difference is greater than 1\%. The regions with large differences are discussed in Fig. \ref{fig:FIGX}.}}
\end{figure}

Figure \ref{fig:FIG7}(a) shows the analyzed sound pressure in the range of $x=[0,l]$ and $t=[0,T]$ after 20,000 training epochs. Although we only analyzed one period, the same result is shown twice along the time axis to confirm the waveform continuity at $t=0$ and $t=T$. The results of the finite difference method (FDM) are shown for comparison in Fig. \ref{fig:FIG7}(b). Figure \ref{fig:FIG7} demonstrates strong agreement between the ResoNet and FDM analysis results. Figure \ref{fig:FIG8} shows the differences between the results of ResoNet and FDM. The difference was less than 1\% in most regions; however, we noted some streaked regions with a difference of 2\% or more. The regions with large differences are discussed later in this section.

\begin{figure}
\includegraphics[width={7cm}]{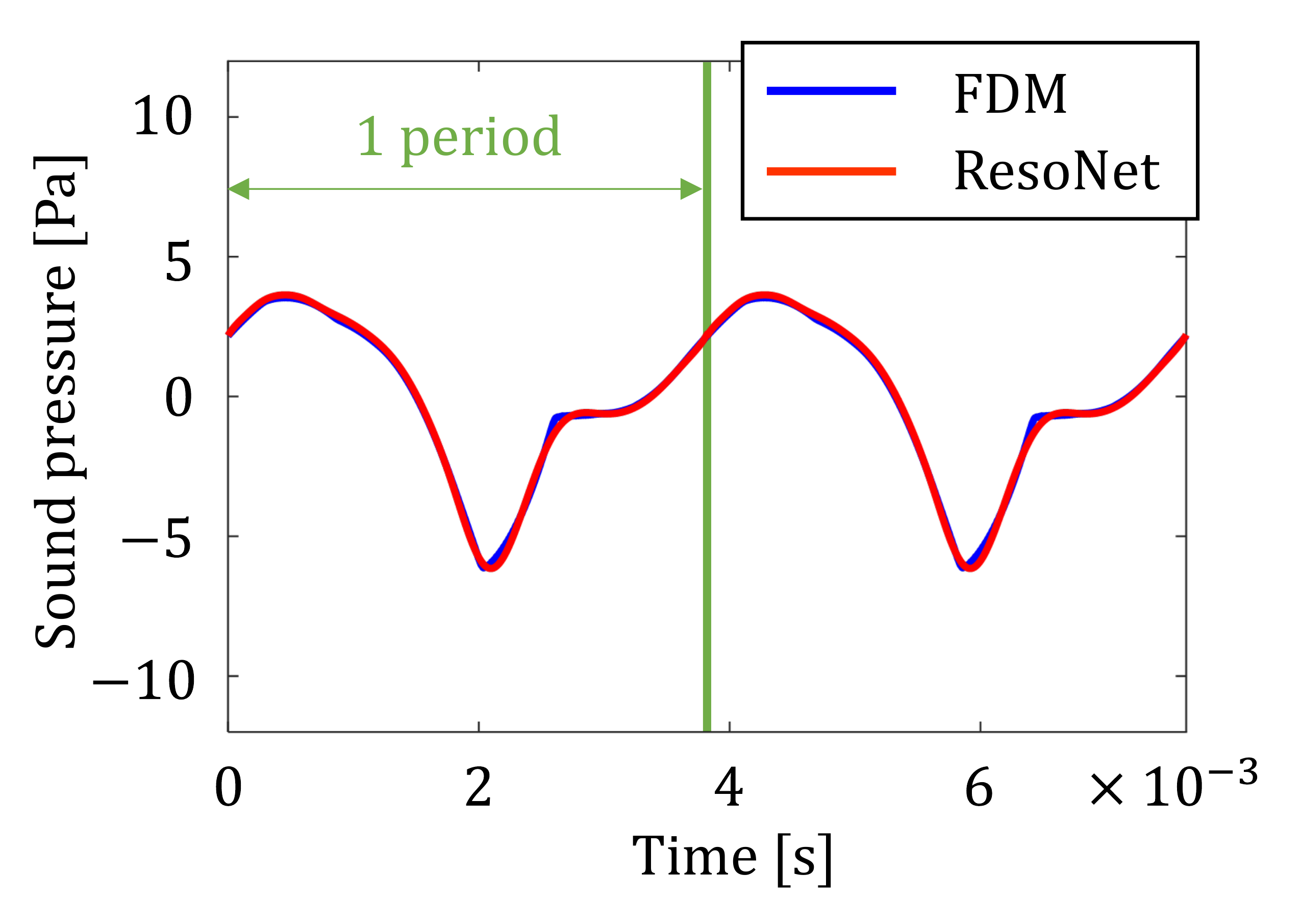}
\caption{\label{fig:FIG9}{(Color online) Sound pressure waveforms at $x=l$. The forward analysis by the proposed method is in good agreement with the results of the finite difference method (FDM), but there are some differences. These differences are discussed in Fig. \ref{fig:FIGX}.}}
\end{figure}

\begin{figure}
\includegraphics[width={7cm}]{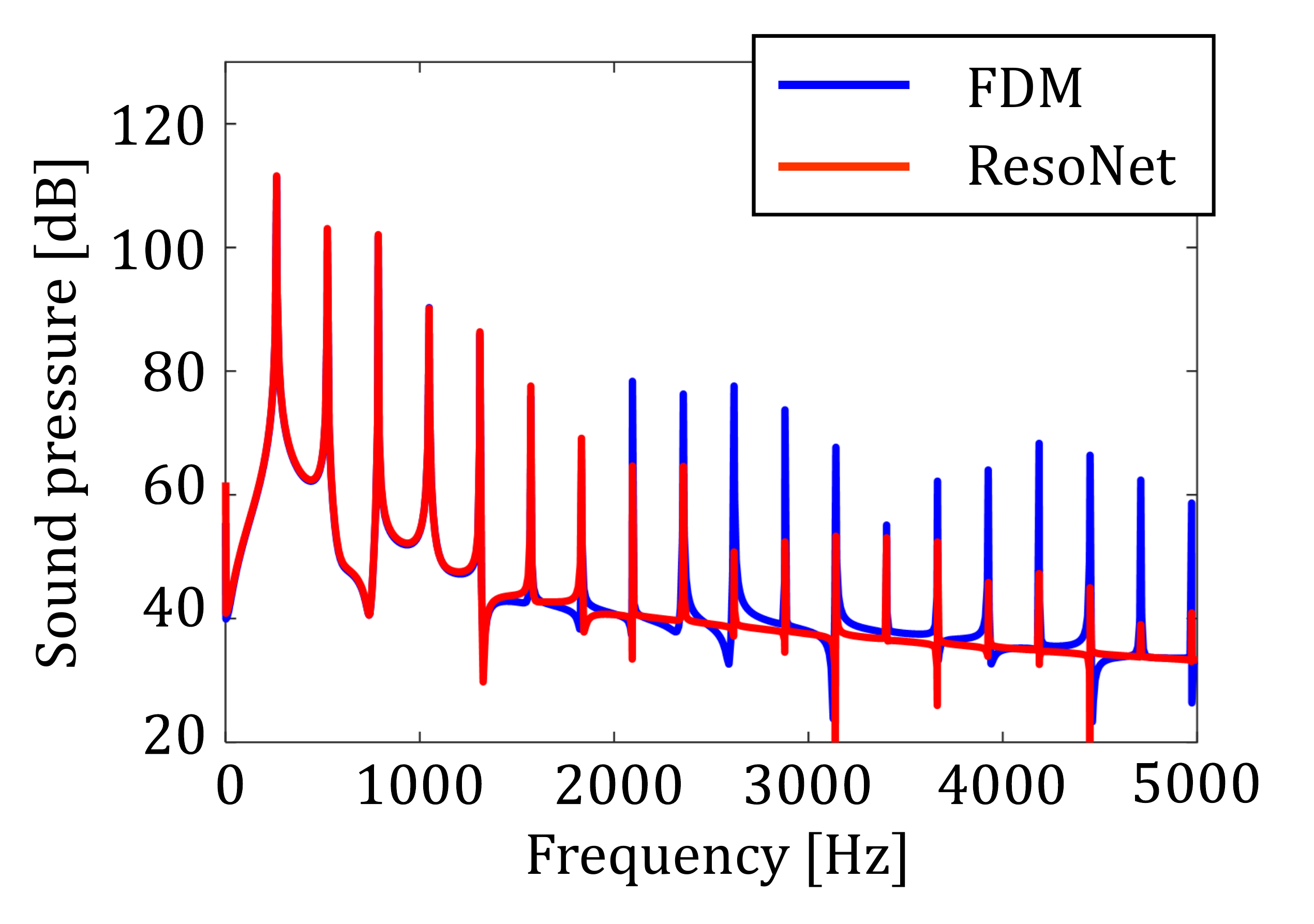}
\caption{\label{fig:FIG10}{(Color online) Frequency spectra of the waveforms of Fig. \ref{fig:FIG9}. The proposed method and FDM agree well up to approximately 2000 Hz, but the differences are larger at higher frequencies. This decrease in neural network approximation accuracy at high frequencies is called F-principle\cite{F-principle}. Fourier features\cite{FF_original} have been proposed as a method to improve the approximation performance at high frequencies, and applying Fourier features in ResoNet is a subject for future study.}}
\end{figure}

Figure \ref{fig:FIG9} shows the analyzed sound pressure waveforms at $x=l$, and Fig. \ref{fig:FIG10} shows the frequency spectra. Although differences are observed in the high-frequency domain in Fig. \ref{fig:FIG10}, the ResoNet results in the time domain indicate its high accuracy in acoustic resonance analysis.

\begin{figure}
\includegraphics[width={7cm}]{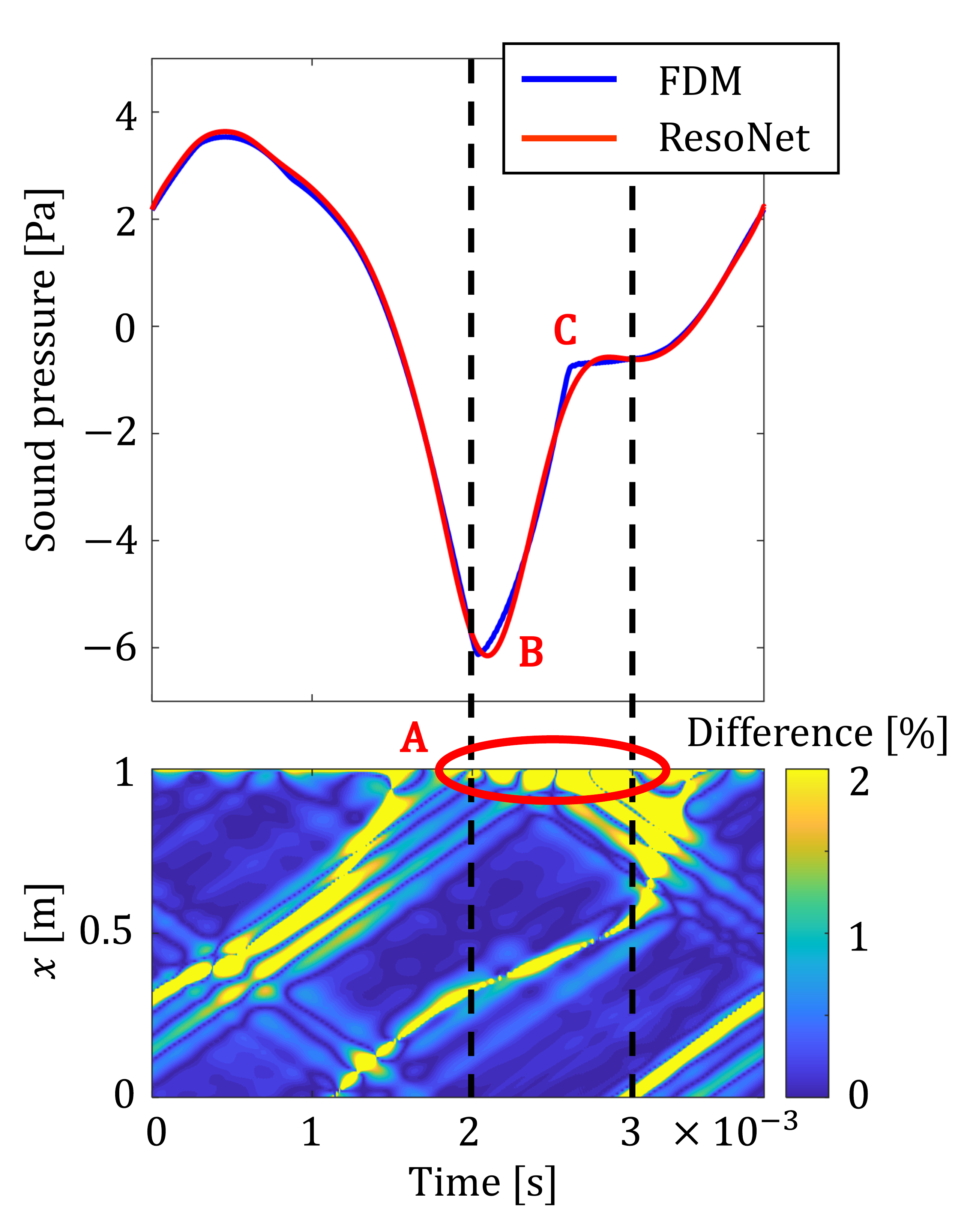}
\caption{\label{fig:FIGX}{(Color online) Sound pressure waveforms at $x=l$ and errors in the same time scale. Due to the F-principle of neural networks\cite{F-principle}, the proposed method does not approximate well the part where the waveform changes steeply. This difference in waveforms leads to the difference in the high-frequency range shown in Fig. \ref{fig:FIG10}.}}
\end{figure}

The regions with large differences in Fig. \ref{fig:FIG8} were discussed considering the sound pressure waveform and frequency spectra. Figure \ref{fig:FIGX} shows the difference and sound pressure waveforms for one period on the same timescale; evidently, the difference is huge in region A. The corresponding region in the sound pressure waveform shows a difference between the FDM and ResoNet waveforms at the points indicated by B and C. At point B, the waveform suddenly changes from monotonically decreasing to monotonically increasing, and at point C, it shifts from monotonically increasing to a horizontal phase. Considering that a neural network is a function approximator, such steep waveform changes may not have been approximated well by ResoNet. This is evident from the difference between FDM and ResoNet in the high-frequency region shown in Fig. \ref{fig:FIG10}. Thus, like other PINNs, ResoNet accurately analyzes in the low-frequency domain; however, its accuracy degrades in the high-frequency domain due to a problem known as the F-principle\cite{F-principle}. Introducing Fourier Features\cite{PINNs_FF} could improve the approximation ability of PINNs at high frequencies, addressing this issue. The problem of reduced approximation accuracy at high frequencies could be improved by modifying the structure of the neural network and learning method, which remain a subject for future studies.

\begin{figure}
\includegraphics[width={9cm}]{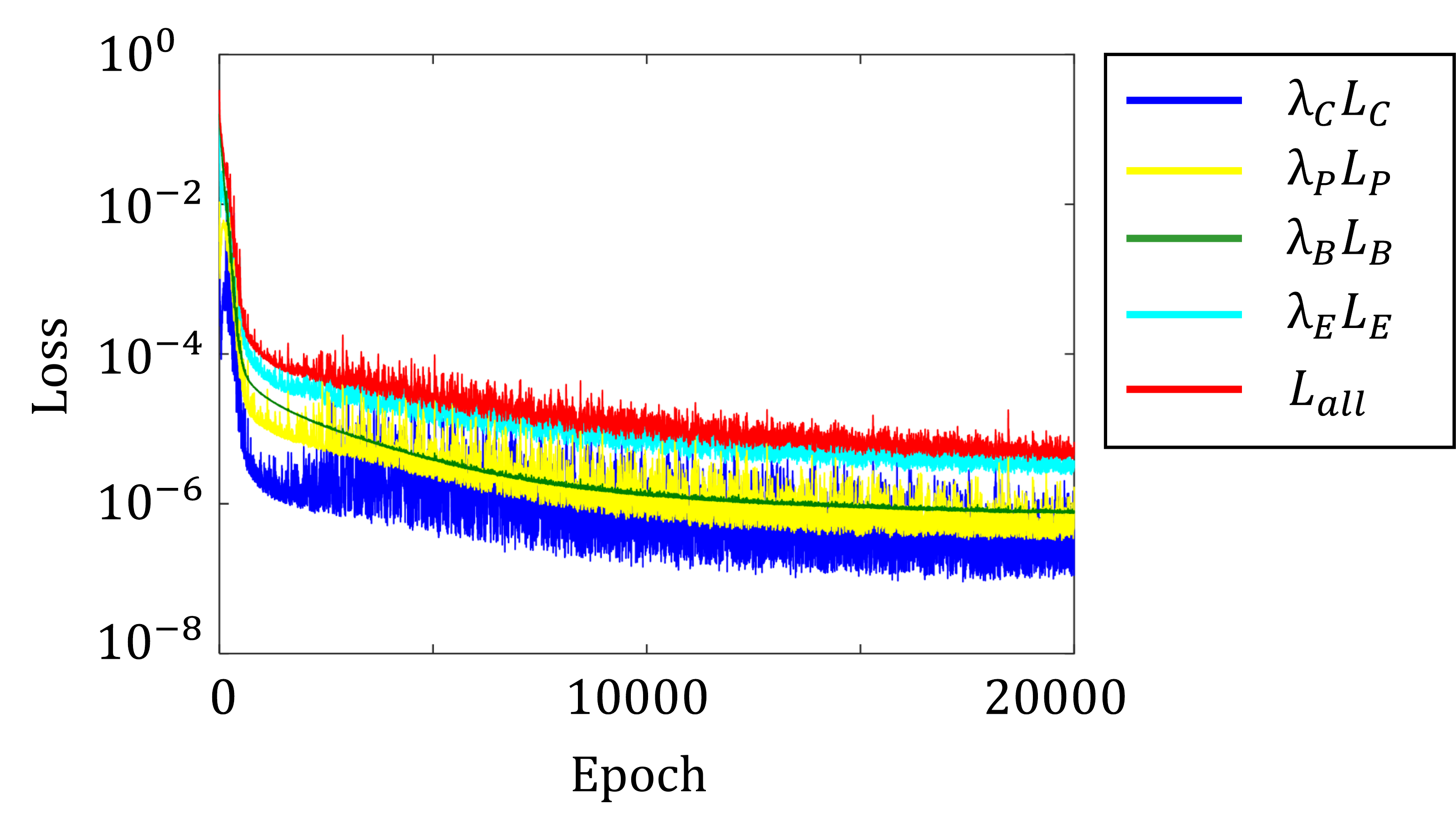}
\caption{\label{fig:FIG12}{(Color online) Losses per epoch. The figure is plotted separately for each loss function. Sufficient convergence is obtained at 20,000 epochs.}}
\end{figure}

To confirm the state of convergence, Fig. \ref{fig:FIG12} shows the loss function for each epoch. Figure \ref{fig:FIG12} shows the total $L_{all}$ and each loss term separately. Since we had confirmed through preliminary simulations that PINN learning stabilizes and generally converges when $L_{all}$ falls below $10^{-5}$, we stopped learning at 20,000 epochs, which fulfills this condition this time.

\subsection{\label{subsec:4:2} Inverse analysis}
Additionally, we performed an inverse analysis on the acoustic tube, as shown in Fig. \ref{fig:FIG1}, using the boundary conditions in Section \ref{subsec:2:2}. The forced-flow waveform and physical properties were identical to those described in Section \ref{subsec:4:1:1}.

We considered two specific situations for inverse analysis: first, the identification of energy loss coefficients, and second, the design optimization of acoustic tubes. The information provided to ResoNet has two waveforms: a flow waveform at $x=0$ and a sound pressure waveform at $x=L$.

\subsubsection{\label{subsec:4:2:1} Additional loss function for inverse analysis}
The loss function for the sound pressure waveform at $x=l$ was introduced into ResoNet using the following procedure: First, the output $\hat{\phi}_{i,M}$ is defined as:
\begin{equation}
\hat{\phi}_{i,M} := F_{w} \left( x_l, t_i; \Theta_{w} \right),
\quad t_i \in \left[ 0,T \right],
\label{Phi_M}
\end{equation}
where $x_l=l$. The loss function for the sound pressure at $x=l$ is defined as:
\begin{equation}
L_{M} = \dfrac{1}{N_M} \sum_{i=1}^{N_M} \left( \hat{p}_{i,M} - \bar{p}_{i,M} \right) ^2,
\label{Loss_M}
\end{equation}
where $N_M$ denotes the number of collocation points for the loss. $\hat{p}_{i,M}$ was obtained from $\hat{\phi}_{i,M}$ by using Eq. (\ref{Calc_p}) and $\bar{p}_{i,M}$, the measured sound pressure waveform, was obtained from the analysis results of FDM simulation. Fig. \ref{fig:FIG12} shows the waveform. The loss function for the entire network is defined as: 
\begin{figure}
\includegraphics[width={7cm}]{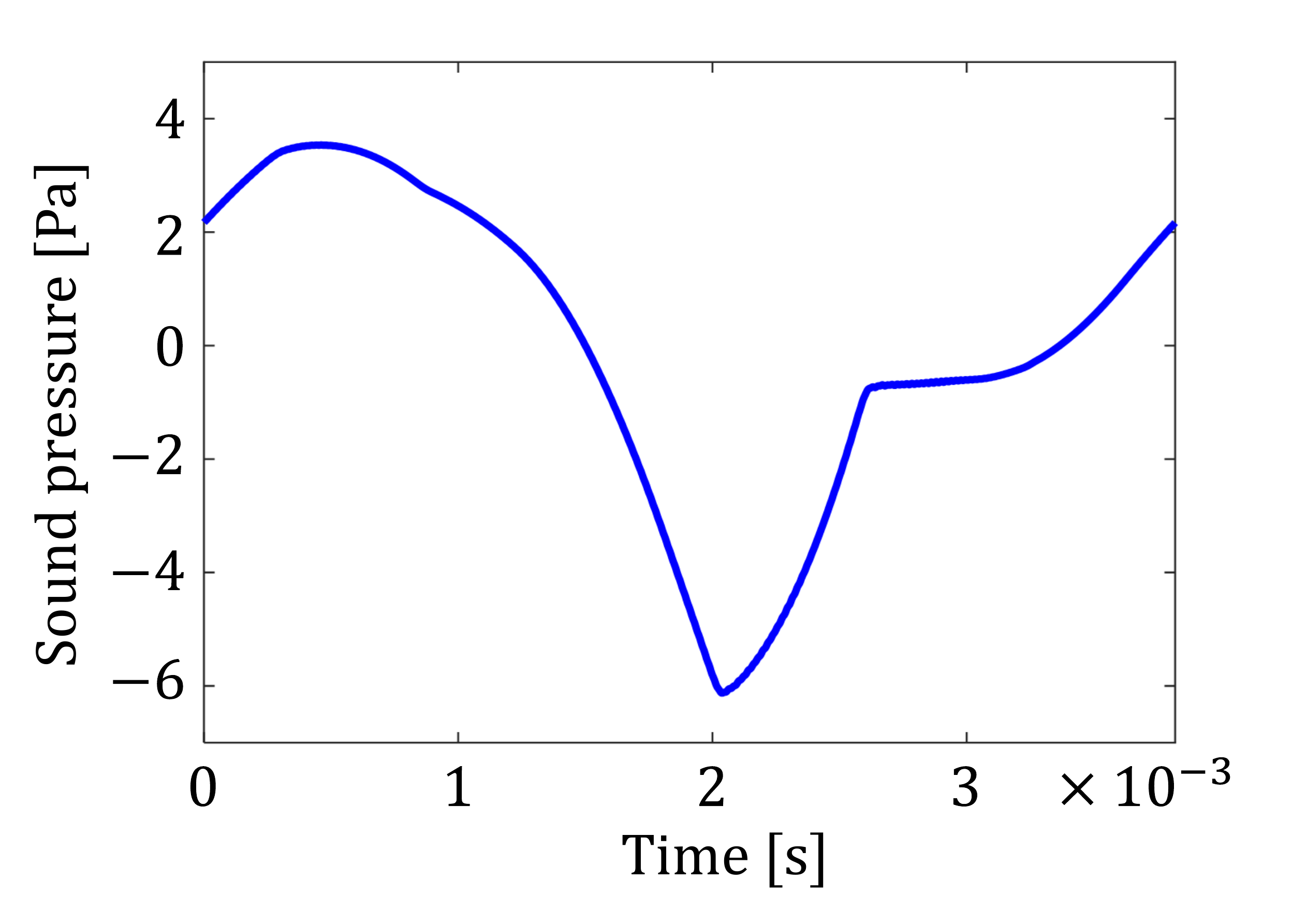}
\caption{\label{fig:FIG13}{(Color online) Sound pressure waveform at $x=l$ (obtained by FDM simulation). This waveform is given to the new loss function $L_M$ as a boundary condition that the neural network's output must satisfy during the inverse analysis.}}
\end{figure}
\begin{equation}
L_{all} = \lambda_E L_E + \lambda_B L_B + \lambda_P L_P + \lambda_C L_C + \lambda_M L_M,
\label{Loss_all_inv}
\end{equation}
where $\lambda_M$ is the weight parameter of $L_M$.

In the inverse analysis, we set the number of nodes in the neural network $N_f$ to 400 and the number of FC blocks $N_b$ to two based on preliminary simulations. We determined the order of the weight parameters of the loss function in the same way as described in Section \ref{subsec:4:1:2}, and the parameters used in the inverse analysis are shown in Table \ref{Table3}. We set the initial value of the learning rate of the Adam optimizer to $\lambda_{\rm{init}}=0.001$ and the decay coefficient to $\beta=0.005$.
\begin{table}[ht]
\caption{\label{Table3}Weight parameters for inverse analysis. We determined these parameters by combining normalization and empirical approaches as in the forward analysis.}
\centering
\begin{ruledtabular}
\begin{tabular}{cccc}
Parameter&Value&Parameter&Value\\
\hline
$\lambda_E$&$5.8$&$\lambda_r$&$1.4 \times 10^{-4}$\\
$\lambda_B$&$1.6 \times 10^8$&$\lambda_u$&$1.6 \times 10^8$\\
$\lambda_P$&$1.0$&$\lambda_p$&$8.7 \times 10^{-6}$\\
$\lambda_C$&$1.0$&$\lambda_t$&$1.3 \times 10^{-12}$\\
$\lambda_l$&$2.9 \times 10^{-6}$&$\lambda_M$&$5.0 \times 10^{-3}$\\
\end{tabular}
\end{ruledtabular}
\end{table}

\subsubsection{\label{subsec:4:2:2} Case 1: Identification of energy loss coefficients}
\begin{figure}
\includegraphics[width={7cm}]{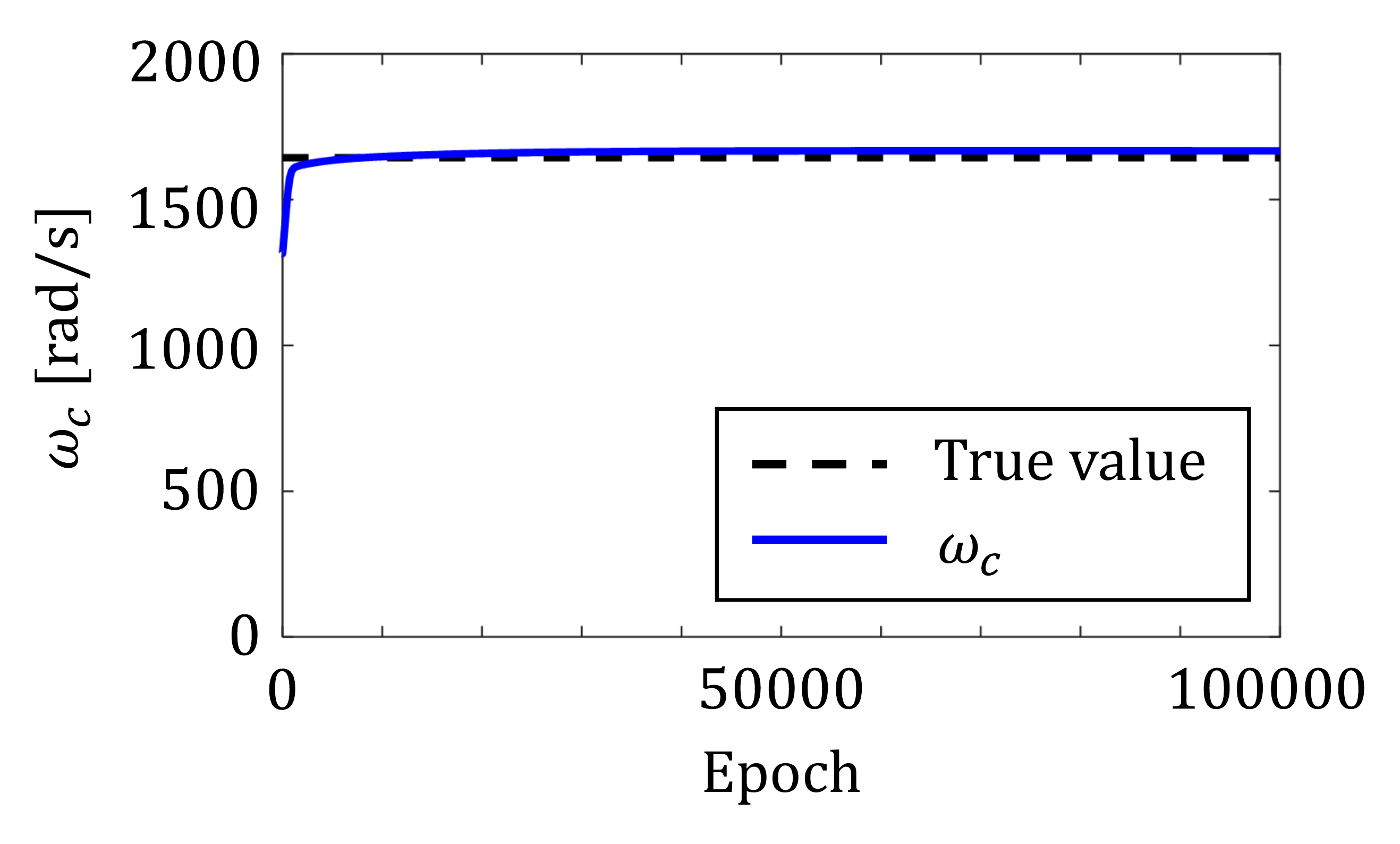}
\caption{\label{fig:FIG14}{(Color online) Identification result of $\omega_c$. An error of 20\% was given as the initial value of $\omega_c$, but the error of $\omega_c$ after 100,000 epochs of training dropped to 1.01\%.}}
\end{figure}

Assuming that the energy-loss coefficients follow Eqs. (\ref{TeleEq_R})--(\ref{TeleEq_G}) and $\omega_c$ in these equations is unknown, we performed an inverse analysis to determine $\omega_c$. As described at the beginning of section \ref{subsec:4:2}, the flow velocity waveform at $x=0$ (Fig. \ref{fig:FIG6}) and the sound pressure waveform at position $x=l$ (Fig. \ref{fig:FIG13}) are given to ResoNet. Because $w_c$ is a trainable parameter of the neural network, we formulated the optimization problem as follows:
\begin{equation}
\min_{\Theta_w,\Theta_r,\omega_c} L_{all}(\Theta_w,\Theta_r,\omega_c).
\end{equation}
Fig. \ref{fig:FIG14} shows the identification results. We set the initial value of $\omega_c$ to $1.3149\times 10^3$ (20\% error) for a true value of $1.6437\times10^3$; however, after 100,000 training epochs, the value converged to $1.6671\times10^3$, which indicated that $\omega_c$ could be identified with an error of 1.01\%.

For reference, we also performed the inverse analysis by mixing 1\% uncorrelated Gaussian noise into $\bar{u}_B$ and $\bar{p}_M$ given as training data. $\omega_c$ after 100,000 epochs of training was 1678.2 (1.02\% error), indicating that the inverse analysis can be performed as well as the noiseless case if the noise is approximately 1\%.

\subsubsection{\label{subsec:4:2:3} Case 2: Design optimization of acoustic tube}
This section describes the design optimization of the length $l$ and diameter $d$ of the acoustic tube that simultaneously satisfies the flow velocity waveform at $x=0$ and the sound pressure waveform at $x=l$. The flow velocity waveform at $x=0$ in Fig. \ref{fig:FIG6} and the sound pressure waveform at $x=l$ in Fig. \ref{fig:FIG12} were given to ResoNet. Because $l$ and $d$ are the trainable parameters of the neural network, the optimization problem was formulated as:
\begin{equation}
\min_{\Theta_w,\Theta_r,l,d} L_{all}(\Theta_w,\Theta_r,l,d).
\end{equation}

The number of collocation points does not change during the calculation, even when encountering variable length. The relative positions of the collocation points at $0 \leq x \leq l$ also do not change during the calculation. We introduced a length-adjusting factor while calculating the partial derivative when handling variable length. Let the initial length of the tube be $l_0$, and the tube length as a variable be $l_v$. The following equation obtains the spatial derivative of $\phi$ at $l=l_v$.
\begin{equation}
\left( \dfrac{\partial \phi}{\partial x} \right)_{l=l_v} =
\dfrac{l_0}{l_v} \left( \dfrac{\partial \phi}{\partial x} \right)_{l=l_0}.
\label{LengthScaling}
\end{equation}
The $\partial \phi / \partial x$ at initial length $l=l_0$ can be obtained by automatic differentiation, and by substituting Eq. (\ref{LengthScaling}) into the partial differential equation when calculating the loss function, it is possible to calculate the loss at $l=l_v$.

\begin{figure}
\includegraphics[width={7cm}]{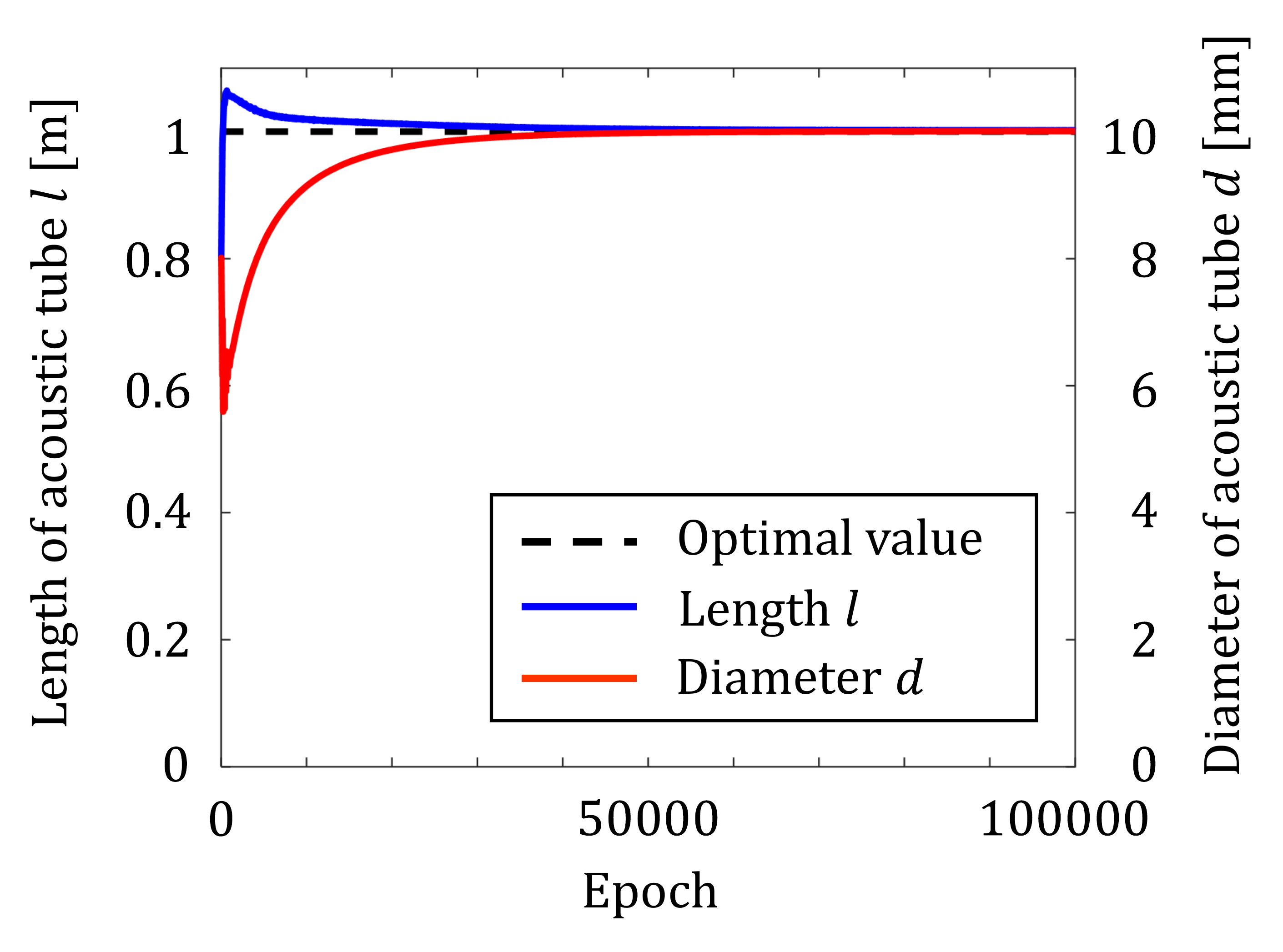}
\caption{\label{fig:FIG15}{(Color online) Identification result of $l$ and $d$. A 20\% error was given as the initial value of $l$ and $d$, but after 100,000 training epochs, they converged to a value close to the optimal value.}}
\end{figure}

Fig. \ref{fig:FIG15} shows the identification results. We set the initial value of $l$ to 0.8 (20\% error) for an optimal value of 1 and the initial value of $d$ to 8 (20\% error) for an optimal value of 10. Table \ref{Table4} indicates that $l$ and $d$ were identified with high accuracy with respect to the optimal values after 100,000 training epochs. Table \ref{Table4} also shows the results when 1\% uncorrelated Gaussian noise was mixed into the training data $\bar{u}_B$ and $\bar{p}_M$ for reference, indicating that the inverse analysis can be performed with the same level of accuracy as the noiseless case.
\begin{table}[ht]
\caption{\label{Table4}Optimal and identified value of $l$ and $d$. An error of 20\% was given as the initial value, but after 100,000 epochs of training, the error was significantly reduced.}
\centering
\begin{ruledtabular}
\begin{tabular}{ccc}
&Length $l$ [m]&Diameter $d$ [mm]\\
\hline
 Optimal&1&10\\
\hline
 Identified (no noise)&1.0022 &10.009\\
 &(0.22\% error)&(0.09\% error)\\
 \hline
 Identified (1\% noise)&1.0024&10.004\\
 &(0.24\% error)&(0.04\% error)\\
\end{tabular}
\end{ruledtabular}
\end{table}

\section{\label{sec:5} Conclusion}
In this study, we proposed ResoNet, a PINN for analyzing acoustic resonance, and demonstrated its effectiveness by performing a time-domain analysis of acoustic resonance by introducing a loss function for periodicity into a neural network.

The forward analysis performed using an acoustic tube of 1 m, the scale of a musical instrument or car muffler, revealed that acoustic resonance analysis could be performed with sufficient accuracy in the time domain. The accuracy of the analysis decreased with abrupt changes in the sound pressure waveform and the high-frequency region in the frequency domain. Given that this is due to the function approximation capability of the neural network, designing a PINN structure that is more suitable for acoustic analysis is a topic for future studies.

We identified the energy loss coefficient in the acoustic tube for the inverse analysis and optimized its design. In these inverse problems, the true and optimal values could be identified with high accuracy from the waveform data at the endpoints of the acoustic tube. 

These results indicate that PINN may be a new option for acoustic inverse problems. This highlights the potential for broad applicability for parameter identification and other inverse problems related to 1D acoustic tubes, such as the design optimization of musical instruments and glottal inverse filtering (GIF). In future work, we intend to address these acoustic inverse problems using ResoNet.

\begin{acknowledgments}
JSPS KAKENHI Grant Number JP22K14447 supported this work.
\end{acknowledgments}

\section*{AUTHOR DECLARATIONS}
\subsubsection*{Conflict of Interest}
The authors have no conflicts to disclose.

\section*{DATA AVAILABILITY}
The data that support the findings of this study are available within the article.

% \appendix
% \section{Derivation of wave equation with energy loss terms [Eq. (\ref{WaveEq})]}
\appendix
\section{Derivation of wave equation}\label{Ap_WaveEq}
This section describes the derivation of the wave equation using the energy-loss terms (Eq. (\ref{WaveEq})). First, the telegrapher's equation in the time domain (Eq. (\ref{WaveEq_Time1}) and (\ref{WaveEq_Time2})).
\begin{align}
\dfrac{\partial u}{\partial x} &= -Gp - \dfrac{A}{K}\dfrac{\partial p}{\partial t}, \label{AP_WaveEq_Time1}\\
\dfrac{\partial p}{\partial x} &= -Ru - \dfrac{\rho}{A}\dfrac{\partial u}{\partial t}. \label{AP_WaveEq_Time2}
\end{align}
The velocity potential is defined as:
\begin{align}
u &= -A\dfrac{\partial \phi}{\partial x}. \label{AP_VelPot1}
\end{align}
By substituting Eq. (\ref{AP_VelPot1}) into Eq. (\ref{AP_WaveEq_Time2}), we obtain:
\begin{align}
p &= RA\phi + \rho\dfrac{\partial \phi}{\partial t}. \label{AP_VelPot2}
\end{align}
By substituting Eqs. (\ref{AP_VelPot1}) and (\ref{AP_VelPot2}) into Eq. (\ref{AP_WaveEq_Time1}), we obtain:
\begin{equation}
\begin{split}
-\dfrac{\partial}{\partial x} \left(A \dfrac{\partial \phi}{\partial x} \right)
&= -G \left(RA \phi + \rho \dfrac{\partial \phi}{\partial t} \right) \\
& \quad - \dfrac{A}{K} \dfrac{\partial}{\partial t} \left(RA \phi + \rho \dfrac{\partial \phi}{\partial t} \right).
\end{split}
\label{AP_WaveEq_before}
\end{equation}
We obtain the following wave equation with energy loss terms by expanding and simplifying Eq. (\ref{AP_WaveEq_before}).
\begin{equation}
\dfrac{\partial ^{2}\phi }{\partial x^{2}}+\dfrac{1}{A}\dfrac{\partial A}{\partial x}\dfrac{\partial \phi }{\partial x}=GR\phi +\left( \dfrac{G\rho }{A}+\dfrac{RA}{K}\right) \dfrac{\partial \phi }{\partial t}+\dfrac{\rho }{K}\dfrac{\partial ^{2}\phi }{\partial t^{2}}.
\label{AP_WaveEq}
\end{equation}

\section{Activation function}\label{Ap_Activation}
\begin{figure}
\includegraphics[width={9cm}]{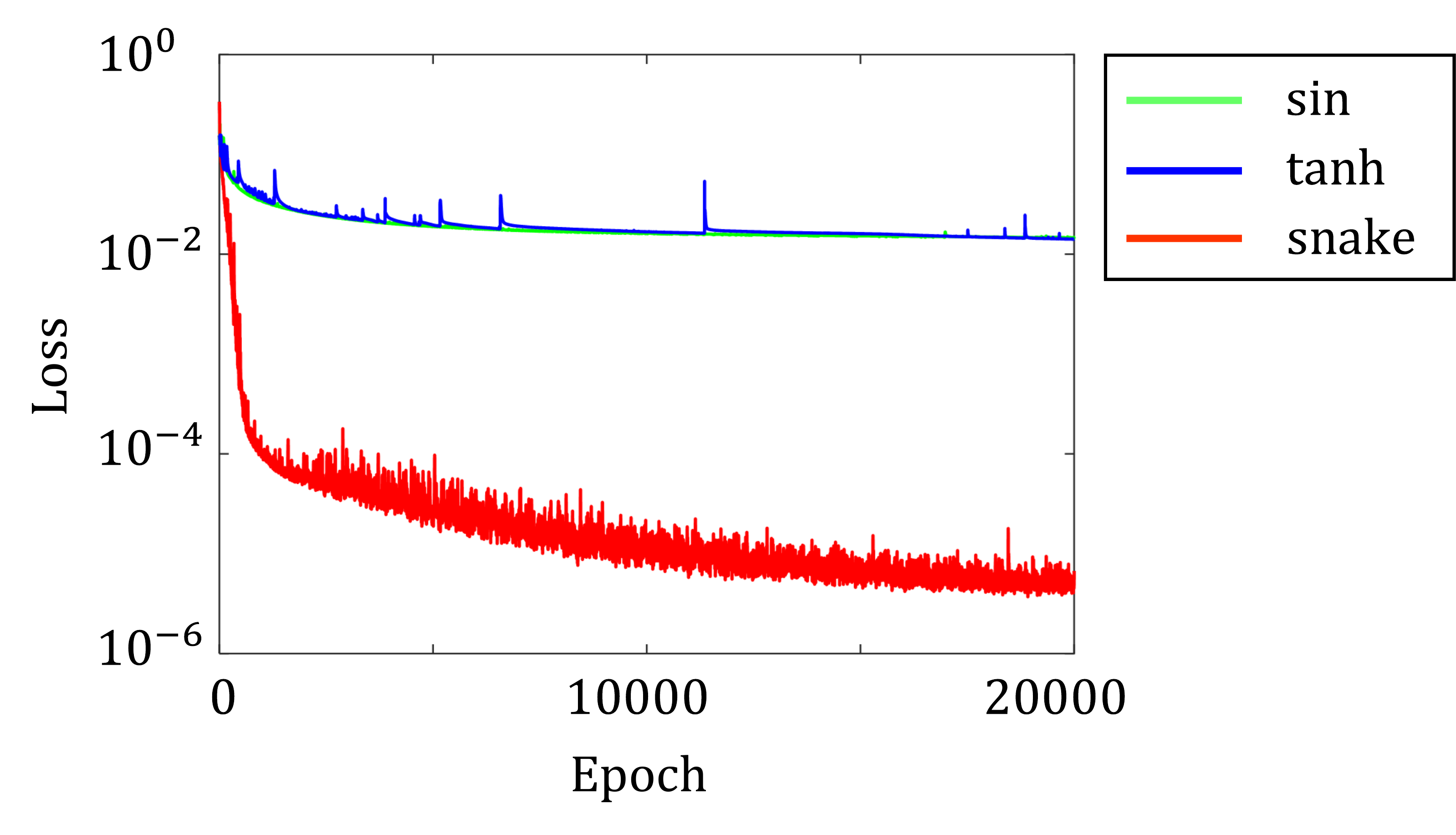}
\caption{\label{fig:FIG16}{(Color online) Comparison of losses for different activation functions. Compared to sin and tanh, snake shows very high convergence performance. The exact reason for this difference in convergence performance is unclear, but snake has a linear term in addition to the periodic nonlinear term, which may have affected the difference in convergence performance.}}
\end{figure}

This section describes ResoNet's choice of activation function. The reason for using snake as the activation function is the better learning performance compared to the more common tanh and sin in preliminary simulations. Figure \ref{fig:FIG16} shows the losses when forward analysis uses tanh, sin, or snake as the activation function.
From the figure, the highest performance is achieved when snake is used as the activation function.

The reason why snake shows the highest performance is not exactly clear. However, some insight can be obtained from the snake equation. The following is a restatement of the snake's equation in the paper.
\begin{equation}
f(a) = a+\sin ^2 a,
\label{AP_EQ_Activation}
\end{equation}
From Eq. (\ref{AP_EQ_Activation}), snake has a linear term in the first term and a periodic $\sin^2a$ in the second term. A periodic activation function such as sin can be regarded as a representation of the signal using the Fourier series \cite{PINN_Borrel_1D,SinFourier}. In the second term, snake with periodic $\sin^2a$ effectively approximates periodic functions \cite{3_Snake}. For these reasons, snake may have been suitable for the periodic resonance state sound waves treated in this study.

Also, unlike sin, snake has a linear term in the first term. Therefore, the vanishing gradient problem is mitigated by using snake compared to tanh and sin, and the learning process could have been more effective at deeper layers by using snake.

\section{Collocation points}\label{Ap_Collocation}
This section describes the validity of the number of collocation points. In the forward analysis, The number of collocation points $N_E$ of the PDE loss is 5000, and the average number of divisions along the time axis is 70.7. Considering that the period of the Rosenberg wave used in this study is $3.82 \times 10^{-3}$ s, this number of divisions corresponds to a sampling frequency of 18.5 kHz. Since the fundamental frequency of the Rosenberg wave is 261.6 Hz, the time resolution of the $N_E$ is sufficient considering the Nyquist theorem. The average number of spatial divisions is similarly 70.7, corresponding to $\Delta x = 0.014$ m when the acoustic tube length $l=1$ m is considered. Assuming that the frequency under analysis is the highest 5000 Hz, the number of sampling points per wavelength $ppw$ at 5000 Hz is
\begin{equation}
ppw = \frac{c}{5000 \Delta x} = 4.81
\label{AP_ppw}
\end{equation}
which satisfies the Nyquist theorem.

The number of collocation points $N_B$ and $N_C$ for BC loss and coupling loss is 1000 in the time direction, which is sufficient from the above discussion. The number of collocation points $N_P$ for Periodicity loss is 1000 in the spatial direction, which is also sufficient from the above discussion.

\begin{figure}
\includegraphics[width={7cm}]{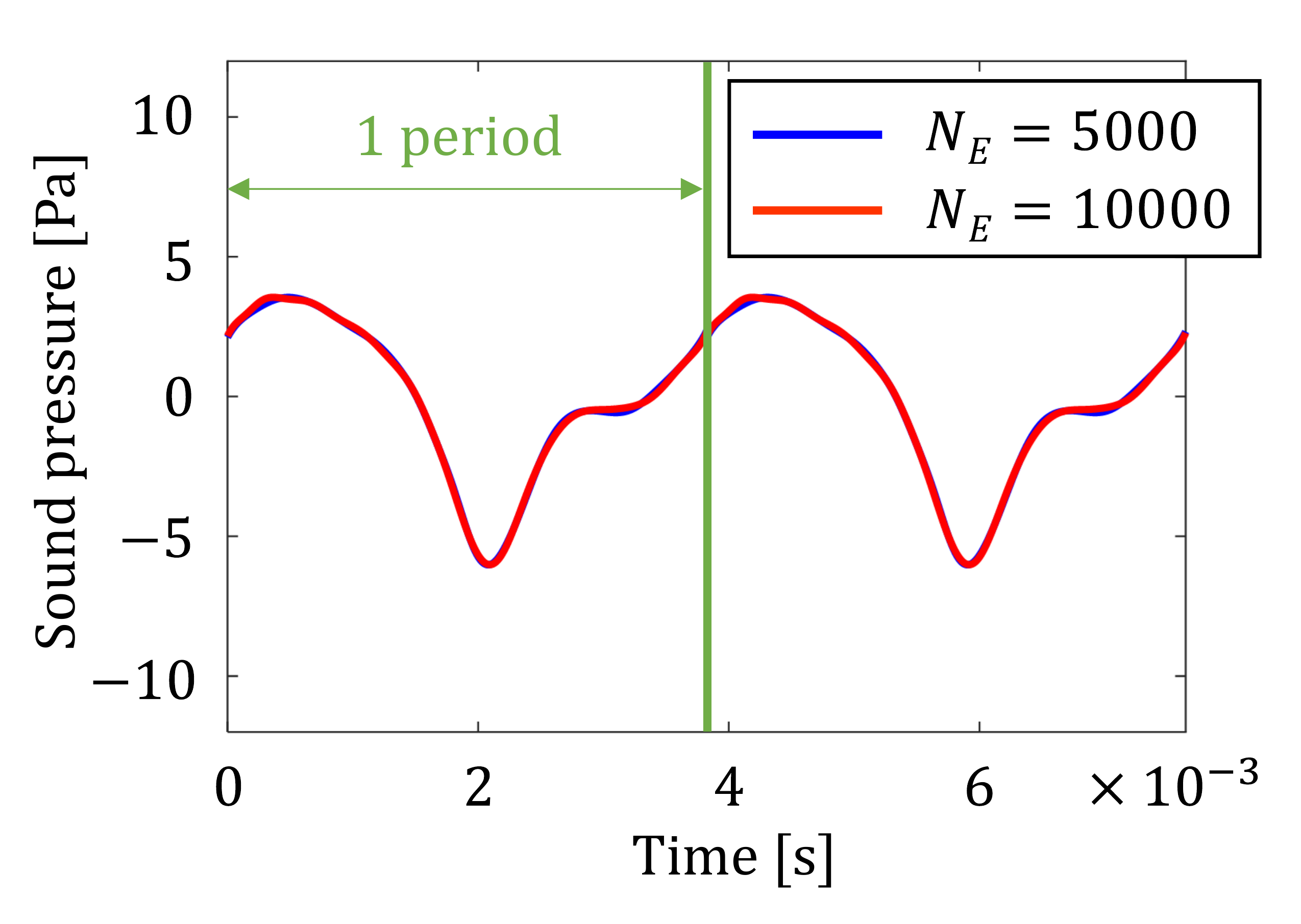}
\caption{\label{fig:FIG17}{(Color online) Sound pressure waveforms at $x=l$ analyzed by ResoNet for different number of collocation points $N_E$. There is little difference between the cases of 5,000 and 10,000 collocation points.}}
\end{figure}

For reference, Fig. \ref{fig:FIG17} compares the case with $N_E$ of 5000 (the number of collocation points adopted in this study) and the case with 10000 and shows that the analysis results with $N_E$ of 5000 are almost similar to those with 10000.

\section{Step Size for FDM}\label{Ap_FDM}
\begin{figure}
\includegraphics[width={7cm}]{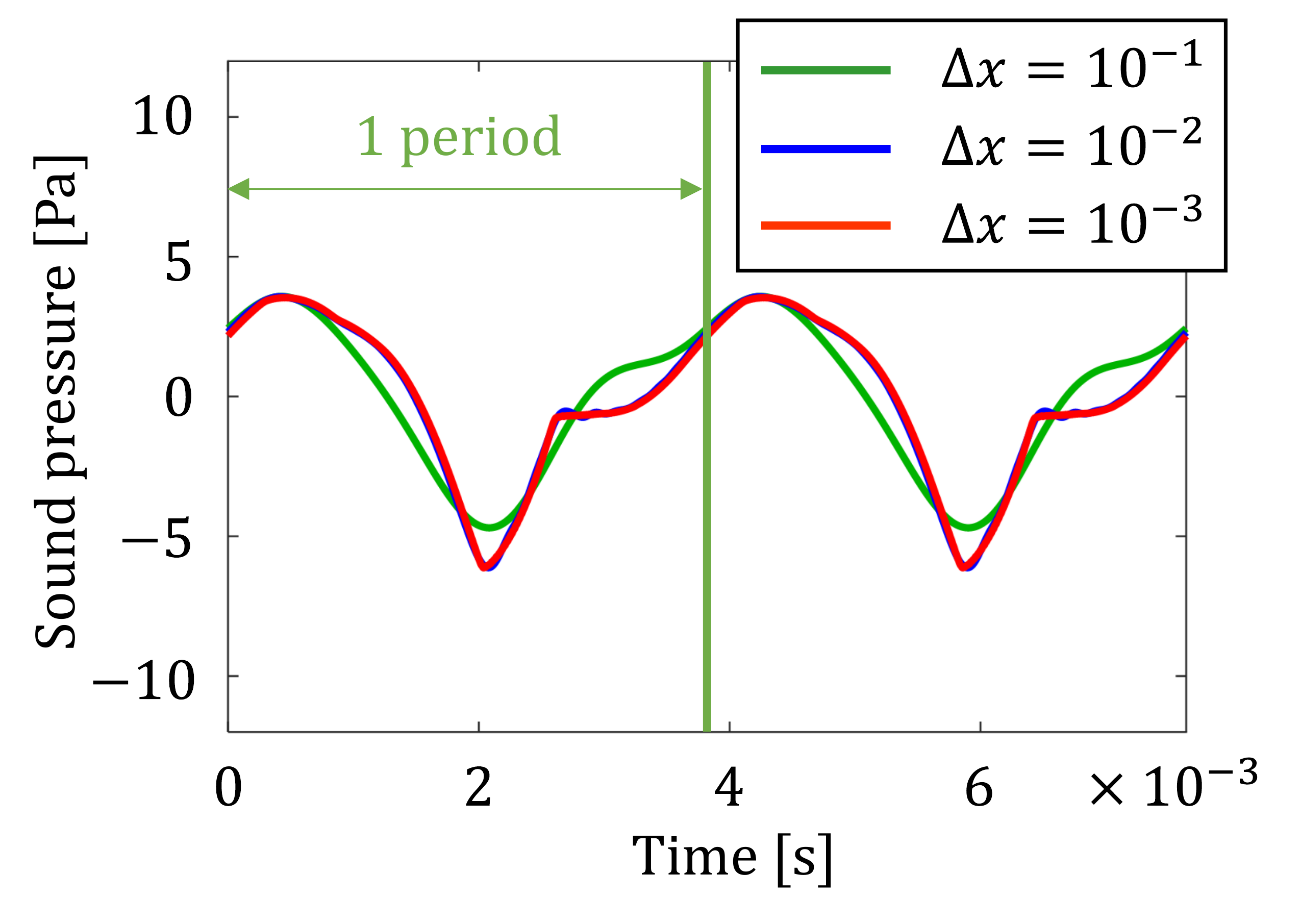}
\caption{\label{fig:FIG18}{(Color online) Sound pressure waveforms at $x=l$ analyzed by finite difference method for different spatial resolutions. Since there is little difference in spatial resolution between the $\Delta x=10^{-2}$ and $\Delta x=10^{-3}$ cases, the $\Delta x=10^{-3}$ used in this study has sufficient resolution.}}
\end{figure}

In this section, we describe the step size of the finite difference method. The time step size of $\Delta t=0.5 \times 10^{-6}$ s corresponds to a resolution of $2 \times 10^6$ Hz, which is sufficient considering that the fundamental frequency of the Rosenberg wave analyzed in this study is 261.6 Hz. For the spatial resolution, a comparison of $\Delta x=10^{-1}$, $10^{-2}$, and $10^{-3}$ m is shown in Fig. \ref{fig:FIG18}. $\Delta x=10^{-2}$ and $10^{-3}$ are almost similar, indicating that $\Delta x=10^{-3}$ is sufficient for the required resolution.

\section{Initial conditions}\label{Ap_Initial}
In this section, we investigate the effect of initial conditions on the inverse analysis. We performed the inverse analysis using the training data, shifted by $T/2$ ($T$: period) from the original $\bar{u}_B$ and $\bar{p}_M$.

The results of the inverse analysis are shown in Table \ref{Table_AP_Init_Omega} and \ref{Table_AP_Init_Design}. Table A shows the result of the identification of $\omega_c$ (Section \ref{subsec:4:2:2}), and Table A shows the result of design optimization (Section \ref{subsec:4:2:3}). Table \ref{Table_AP_Init_Omega} and \ref{Table_AP_Init_Design} show that, although there are some differences in accuracy, the inverse analysis can be performed with accuracy close to that of the original conditions, even when the data given as initial conditions are different.

\begin{table}[ht]
\caption{\label{Table_AP_Init_Omega}Optimal and identified value of $\omega_c$. Even with different initial conditions, the proposed method can perform the inverse analysis with comparable accuracy.}
\centering
\begin{ruledtabular}
\begin{tabular}{cc}
&$\omega_c$ [rad/s]\\
\hline
 Optimal&1643.7\\
\hline
 Identified (original)&1667.1 (1.01\% error)\\
 \hline
 Identified ($T/2$ shifted)&1660.4 (1.01\% error)\\
\end{tabular}
\end{ruledtabular}
\end{table}

\begin{table}[ht]
\caption{\label{Table_AP_Init_Design}Optimal and identified value of $l$ and $d$. Even with different initial conditions, the proposed method can perform the inverse analysis with comparable accuracy.}
\centering
\begin{ruledtabular}
\begin{tabular}{ccc}
&Length $l$ [m]&Diameter $d$ [mm]\\
\hline
 Optimal&1&10\\
\hline
 Identified (original)&1.0022 &10.009\\
 &(0.22\% error)&(0.09\% error)\\
 \hline
 Identified ($T/2$ shifted)&1.0031&9.8610\\
 &(0.31\% error)&(1.39\% error)\\
\end{tabular}
\end{ruledtabular}
\end{table}

\bibliography{Bibliography}

\end{document}